\documentclass[iop]{emulateapj}
\usepackage{verbatim} 
\usepackage{ulem}

\begin{document}

\title{First Detection of A Sub-kpc Scale Molecular Outflow in the Starburst Galaxy NGC 3628}

\author{
       An-Li Tsai\altaffilmark{1,2,3},
       Satoki Matsushita\altaffilmark{2},
       Albert K. H. Kong\altaffilmark{4},
       Hironori Matsumoto\altaffilmark{5},
       Kotaro Kohno\altaffilmark{6,7},
\\altsai@astro.ncu.edu.tw}

\affil{$^1$ Graduate Institute of Astronomy, National Central University, 
 No. 300, Jhongda Rd., Jhongli, Taoyuan 32001, Taiwan}

\affil{$^2$ Institute of Astronomy and Astrophysics, Academia Sinica, 
 P.O. Box 23-141, Taipei 10617, Taiwan}

\affil{$^3$ Department of Earth Sciences, National Taiwan Normal University, 
No. 88, Sec. 4, Tingzhou Rd., Taipei 11677, Taiwan}

\affil{$^4$ Institute of Astronomy and Department of Physics, National Tsing Hua University,
 No. 101, Sec. 2, Kuang-Fu Rd., Hsinchu 30013, Taiwan}
\affil{$^5$ Department of Physics, Nagoya University, 
  Furo-cho, Chikusa-ku, Nagoya, Aichi 464-8602, Japan}
\affil{$^6$ Institute of Astronomy, University of Tokyo, 
 2-21-1 Osawa, Mitaka, Tokyo 181-0015, Japan}
\affil{$^7$ Research Center for the Early Universe, University of Tokyo, 7-3-1 Hongo, Bunkyo, Tokyo 113-0033, Japan}

\begin{abstract}
We successfully detected a molecular outflow with a scale of 370 -- 450~pc
in the central region of the starburst galaxy NGC~3628 through
 deep CO(1-0) observations by using the Nobeyama Millimeter Array (NMA).
The mass of the outflowing molecular gas is $\sim$~2.8~$\times$~10$^7$~M$_\sun$,
 and the outflow velocity is $\sim$~90$\pm$10~km~s$^{-1}$.
The  expansion timescale of the outflow is 3.3 -- 6.8~Myr, and 
the molecular gas mass flow rate is 4.1 -- 8.5~M$_{\sun}$~yr$^{-1}$.
It requires mechanical energy of (1.8 -- 2.8)~$\times$~10$^{54}$~erg 
 to create this sub-kpc scale molecular outflow.
In order to understand the evolution of the molecular outflow,
 we compare the physical properties
between the molecular outflow observed from our NMA CO(1-0) data and
 the plasma gas from the soft X-ray emission of
 the Chandra X-ray Observatory (CXO) archival data.
We found that the distribution between the molecular outflow
 and the strong plasma outflow seems to be in a similar region.
In this region, the ram pressure and the thermal pressure of the plasma outflow
 are  10$^{-(8-10)}$ dyne~cm$^{-2}$,
and the thermal pressure of molecular outflow 
is 10$^{-(11-13)}$ dyne~cm$^{-2}$.
This implies the molecular outflow is still expanding outward.
The molecular gas consumption timescale is estimated as 17 -- 27~Myr, 
 and the total starburst timescale is 20 -- 34~Myr.
The evolutionary parameter is 0.11 -- 0.25,
 suggesting that the starburst activity in NGC 3628 is still
 in a young stage.
\end{abstract}


\maketitle
\thispagestyle{empty}

\section{Introduction}
Molecular gas consumption,
 mainly dominated by star formation and losing gas through outflow,
 plays an important role on star formation.
To understand the molecular gas consumption in galaxies
 help us to study galaxies evolution.

Galaxies with violent star formation, especially in central nuclear region,
 are called starburst galaxies.
They have star formation rates (SFRs) several to hundreds times
 higher than normal spiral galaxies.
The strong star formation activities of starburst galaxies 
produce a large  number of young and massive OB stars in a very short time.
These massive stars finally end up their lifetime with supernova explosions.
Therefore, vast amount of stellar winds from massive stars and supernova explosions
generate huge energy and high pressure to create high velocity galactic winds,
which sweep up the surrounding interstellar medium (ISM)
and create galactic scale outflows or superbubbles. 
These scenarios have been described by 
several analytical models (e.g., \citealt{che85,mcc87,yok93}), 
numerical simulations (e.g., \citealt{tom88,coo08}), and
observational results (e.g., \citealt{hec90,cec02}).
Previous observations mainly focused on the warm ionized and hot plasma outflows, 
such as H$\alpha$ emission observations (e.g., \citealt{arm90,leh96})
and soft X-ray observations (e.g., \citealt{str04}).
However, the observations on molecular outflows or superbubbles are very rare in the past.
Lack of this study will cause under-estimation on the molecular gas consumption in galaxies,
and have a bias to understand the galaxies evolution in our universe.

The reasons for the rarity of molecular gas detections were mostly due to their
diffuse/extended nature and poor instrumental sensitivities.
Recent improvements of various instruments 
provide a better chance to study the molecular outflows and superbubbles
in galaxies.
The recent techniques to detect molecular outflows and superbubbles in galaxies
 can be classified as three types:
(1) High-velocity CO wings to detect molecular outflows,
 e.g., NGC~3256 \citep{sak06b}, Mrk~231 \citep{fer10}, 
        M51 \citep{mat07}, NGC~1266 \citep{ala11},
	and local ULIRGs \citep{chu11}.
(2) P-cygni profile to detect molecular superbubbles,
 e.g., Arp~220 \citep{sak09}, and Mrk~231 \citep{fis10}.
(3) Direct imaging at millimeter waveband for
 CO molecular outflows and superbubbles,
 e.g., molecular outflow in M82 \citep{nak87, wal02}, 
       molecular superbubble in M82 \citep{wei99, mat00, mat05}, and
       molecular superbubbles in NGC 253 \citep{sak06}.
The first and second techniques are useful to detect outflows and superbubbles
 from face-on galaxies.
However, it is difficult to distinguish the molecular gas 
 of outflows/superbubbles from that of galactic disk.
Therefore, we cannot have an accurate measurement on 
 molecular gas mass losing.
To solve this problem, we need 
 to observe edge-on galaxies by using the third technique.
Although this technique is time consuming,
 we can directly observe the morphology of molecular outflows and superbubbles,
 and therefore can have an accurate measurement on 
 molecular gas mass losing.

So far, only a few molecular outflows and superbubbles in galaxies have been 
directly imaged by using the third technique.
More observations to directly image molecular outflows and superbubbles
 are necessary
in order to understand the general properties of these structures
 and their influence on starburst activities.
We have been conducting detail studies of 
one molecular outflow and two superbubbles 
toward the nearby edge-on starburst galaxy NGC 2146 \citep{alt09}.
Here, we provide another sample, NGC~3628.

Starburst galaxy NGC~3628 is a member of the interacting Leo Triplet 
(Arp~317), including NGC~3627, NGC~3623, and NGC~3628.
NGC~3628 is the north source in the Leo Triplet, NGC~3627 is the south one,
while NGC~3623 is the southwest one.
Two signatures of tidal interaction in Leo Triplet are observed
in several wavebands:
A plume extending $\sim$~100~kpc from the eastern edge of NGC~3628 
is detected in optical, far-infrared, and H{\small I} emission \citep{chr98};
a bridge between NGC~3628 and NGC~3627 
is detected in H{\small I} emission \citep{col98}.
These structures have been modeled by \citet{rot78} 
and were explained as a tidal interaction
 between NGC~3628 and NGC~3627 about 800~Myr ago.

\begin{figure}
\centering
\epsscale{1}
\plotone{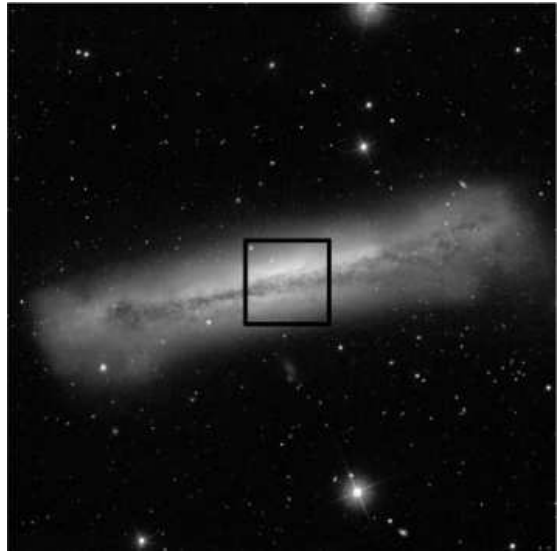}
\caption{
\label{opt-dss}
DSS Optical Image of NGC 3628.
The box size is $2\arcmin\times2\arcmin$,
 the same size of NMA CO(1-0) moment maps showing in Figure~\ref{mnt}.
}
\end{figure}

NGC~3628 is a nearby (D = 7.7~Mpc; 1$\arcsec$ = 37~pc; \citealt{tul88}),
edge-on ({\it i} = 87$\arcdeg$; \citealt{tul88}),
and IR luminous (log $L_{\rm IR}$/L$_{\sun} = 10.25$; \citealt{san03}) galaxy.
The optical image of NGC~3628 looks like a boxy shape (\citealt{chr98};
 see Figure~\ref{opt-dss}).
Radio observations 
 \citep{con82} show that NGC~3628 has a circumnuclear starburst.
Besides, a large-scale galactic wind has been detected in NGC~3628.
Observations with ROSAT \citep{dah96} and 
Chandra X-ray Observatory (CXO) \citep{str01,str04} 
 show that NGC~3628 has an asymmetric plasma bipolar outflow
 with $\sim$ 7 -- 10 kpc scale in both north and south parts;
optical H$\alpha$ observations also show that NGC~3628
 has a $\sim$~10~kpc-scale warm plasma gas outflow \citep{str04}.
It is natural to expect the existence of molecular outflows or superbubbles.
\citet{irw96} claimed that their  
  CO(1-0) Nobeyama Millimeter Array (NMA) observations
detected four expanding molecular superbubbles,
 which are associated with the low velocity gradient ridge outside of the nuclear disk.
However, no molecular outflow has been detected.
Our new CO(1-0) NMA observation detect a sub-kpc scale molecular outflow
 for the first time.
We will discuss its properties,
 its impact on NGC~3628, and its evolution 
 by comparing with the CXO archival data.

\section{Observations and Archive Data}
\subsection{NMA CO(1-0) Observations}
We have used the NMA to observe the CO(1-0) emission 
of NGC~3628 in the central $\sim$~1$\arcmin$ region. 
The observations were made during 2000 December to 2002 February with 3 different configurations of six 10-meter antennas.  
The total on-source time was $\sim$~40~hours.
The phase tracking center was 
$\alpha{\rm (J2000)} = 11^{\rm h}20^{\rm m}17{\fs}011$ and
$\delta{\rm (J2000)} = 13\arcdeg35\arcmin20\farcs064$.
We used tunerless SIS receivers \citep{sun94}, and observed
the CO(1-0) line in the upper side band.
Double side band system temperature was about 200 -- 300~K for most of the
observations.
The backend used was the XF-type spectro-correlator Ultra Wide Band
Correlator (UWBC; \citealt{oku00}), configured to have 512~MHz bandwidth
with 256 channels (i.e., a 2~MHz channel width or 5.2~km~s$^{-1}$ velocity
resolution).
We observed 3C~273 and 3C~279 as bandpass calibrators.
The phase calibrator was 1055+018,
 and amplitude calibrators were 1055+018, 3C~84, and 3C~345.
The flux scale of 1055+018 in 2000 -- 2001 was determined by comparison with Mars.
That in 2001 -- 2002 was determined by comparison with 3C~84 and 3C~345,
 which were determined by comparison with Mars, Uranus, and Neptune.
The uncertainties in the absolute flux scales are
 estimated as  5 -- 6$\%$ in 2000 -- 2001 and  10 -- 12$\%$ in 2001 -- 2002.

The data were calibrated by using the NRO software package ``UVPROC II''
\citep{tsu97}, and were CLEANed by using standard procedures
implemented in the NRAO Astronomical Image Processing System (AIPS).
The maps were made with natural weighting,
with a final synthesized beam
size of $3\farcs01\times2\farcs36$ (147~pc~$\times$~115~pc),
which has a higher  resolution than the past CO(1-0) observations 
($3\farcs93\times2\farcs79$  or 128~pc~$\times$~123~pc;
\citealt{irw96}).
The position angle of the beam is $143\arcdeg$.
The noise level
is 9.86~mJy~beam$^{-1}$ with a velocity resolution of 5.2~km~s$^{-1}$,
better than the previous observation of \citet{irw96},
  65~mJy~beam$^{-1}$ with a velocity resolution of 13~km~s$^{-1}$.

\subsection{CXO Archive Data}

We obtained X-ray data from the CXO archive, 
 which were originally observed by Strickland et al.
 in 2000 December 2 with 60~ks total exposure time.
It was observed with the Advanced CCD Imaging Spectrometer (ACIS). 
The nucleus of NGC~3628 is placed in the ACIS-S3 chip,
 which is a back illuminated CCD on the spectroscopic array.
The back illuminated CCD is more sensitive to low-energy X-ray photons  
 than the front-illuminated ACIS chips \citep{str02,inu05},
 and therefore is powerful to observe diffuse X-ray emission in low energies,
 such as the X-ray superwinds. 
The diffuse gas images of NGC~3628 have been published in \citet{str04},
 but the detail properties have not yet been discussed.
In order to understand the properties of the diffuse gas,
 such as density, mass, energy, and pressure,
 we reprocessed the data, extracted the spectra, and performed model fittings.

The data were reprocessed by using
 the Chandra Interactive Analysis of Observations software package (CIAO) version 4.2,
 and the Chandra Calibration Database (CALDB) version 4.2.2, released on 19 April 2010.
We used
 the High Energy Analysis software (HEAsoft) version 6.5,
 and the X-Ray Spectral Fitting Package (XSPEC) version 12.4
 for further analysis.
The background is chosen from a source-free circular region
 with a radius of 1$\farcm$3 centered at the position
$\alpha{\rm (J2000)} = 11^{\rm h}20^{\rm m}04{\fs}86$ and
$\delta{\rm (J2000)} = 13\arcdeg30\arcmin50\farcs49$,
$\sim5\arcmin$ away from the galactic center.
The background level was 0.04 counts~s$^{-1}$~arcmin$^{-2}$.
To avoid the bias of choosing background, 
 we double checked with blank-sky subtraction.
We used CIAO script DEFLARE and chose {\it sigma} as the flare-cleaning method.
After clipping data with count rates less than 3$\sigma$,
 the exposure time with good time intervals turned to be 56,319~s.

We only consider the data with energy between 0.3~keV and 7.0~keV.
This is because the data with energy above 7.0~keV is dominated by X-ray background,
 and the data quality below 0.3~keV has large uncertainty due to poor calibration.
We grouped the energy bins of the spectra from the selected region
 so that each bin would contain at least 20 counts
to allow $\chi^2$ statistics for spectral fitting.
In order to analyze the diffuse and extended X-ray emission,
 we used two CIAO threads, WAVDETECT and CELLDETECT,
 to detect point sources and removed them from the X-ray data.
After removing detected point sources from the X-ray data,
 we used the CIAO thread DMFILTH to interpolate the emission
 at point source regions from their surrounding backgrounds.

\section{Results}

\subsection{NMA CO(1-0) data}
\subsubsection{Distribution}
\label{n3628-overall-prop}
\begin{figure*}
\centering
\epsscale{1}
\plotone{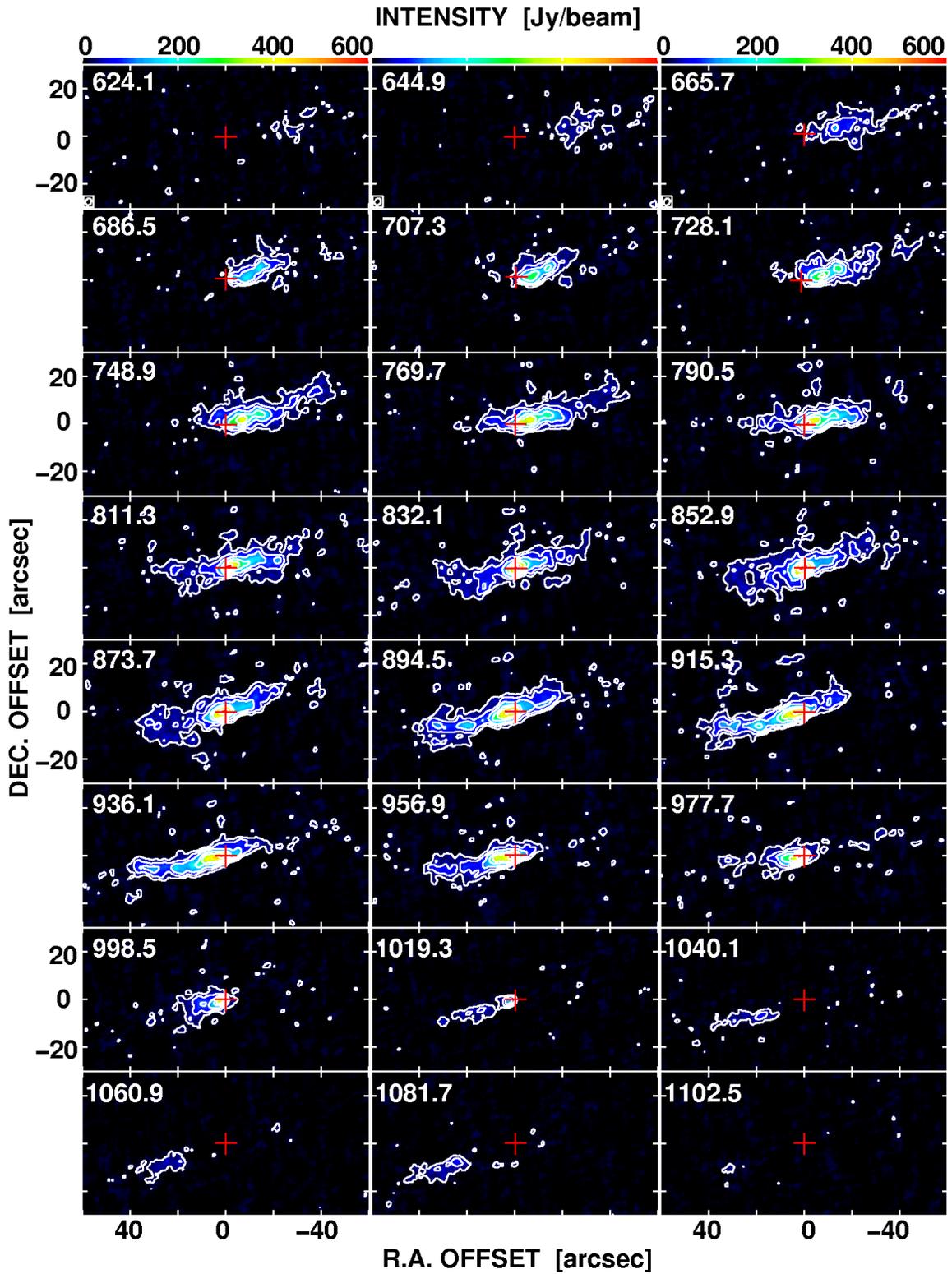}
\caption{
\label{chan}
NMA CO(1-0) channel maps of NGC 3628.
The red cross is the phase tracking center of our observation,
$\alpha{\rm (J2000)} = 11^{\rm h}20^{\rm m}17{\fs}011$ and
$\delta{\rm (J2000)} = 13\arcdeg35\arcmin20\farcs064$.
The synthesize beam size is  $3\farcs01 \times 2\farcs36$  (112.4~pc $\times$ 88.1~pc),
which is shown in the lower left corner of the first panel.
The contour levels are 4, 10, 20, 40, 60, 100, and 130$\sigma$,
where 1$\sigma$ is 4.93~mJy~beam$^{-1}$.
The LSR velocity, in units of km~s$^{-1}$,
 is showing at the top left corner of each channel map.
}
\end{figure*}

We summed every 4 channels (= 20.8~km~s$^{-1}$) from the original channel maps
 to reduce the rms noise level.
The CO(1-0) channel maps are shown in Figure~\ref{chan}.
The noise level is 4.93~mJy~beam$^{-1}$ (68.8~mK).
 Comparing to the noise level of \citet{irw96}, 
 65~mJy~beam$^{-1}$ with a velocity resolution of 13~km~s$^{-1}$,
 which corresponds to 52~mJy~beam$^{-1}$ with a velocity resolution of 20.8~km~s$^{-1}$,
the emission that \citet{irw96} detected is $\sim$ 10$\sigma$ and the second contour in our Figure~\ref{chan}.
This suggests that we could detect 10 times weaker emission than that of \citet{irw96}.
The emission in each channel shows that most of the gas is distributed along the galactic disk,
and the structure in either one or both edges of the galactic disk is fragmented.
Besides, some extended diffuse features
appear in the north of the galactic center in several channels.

\begin{figure*}
\centering
\epsscale{0.99}
\plottwo{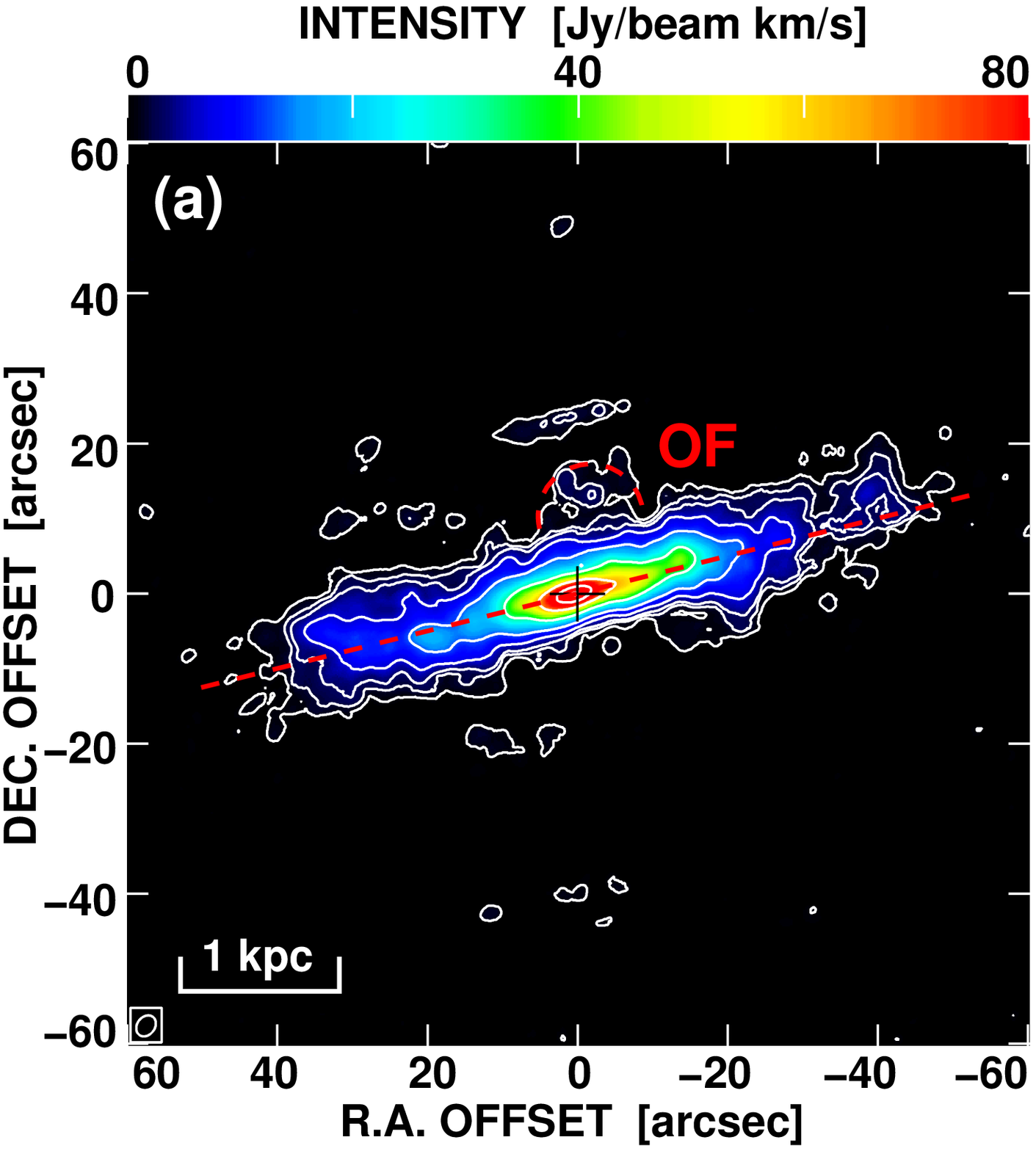}{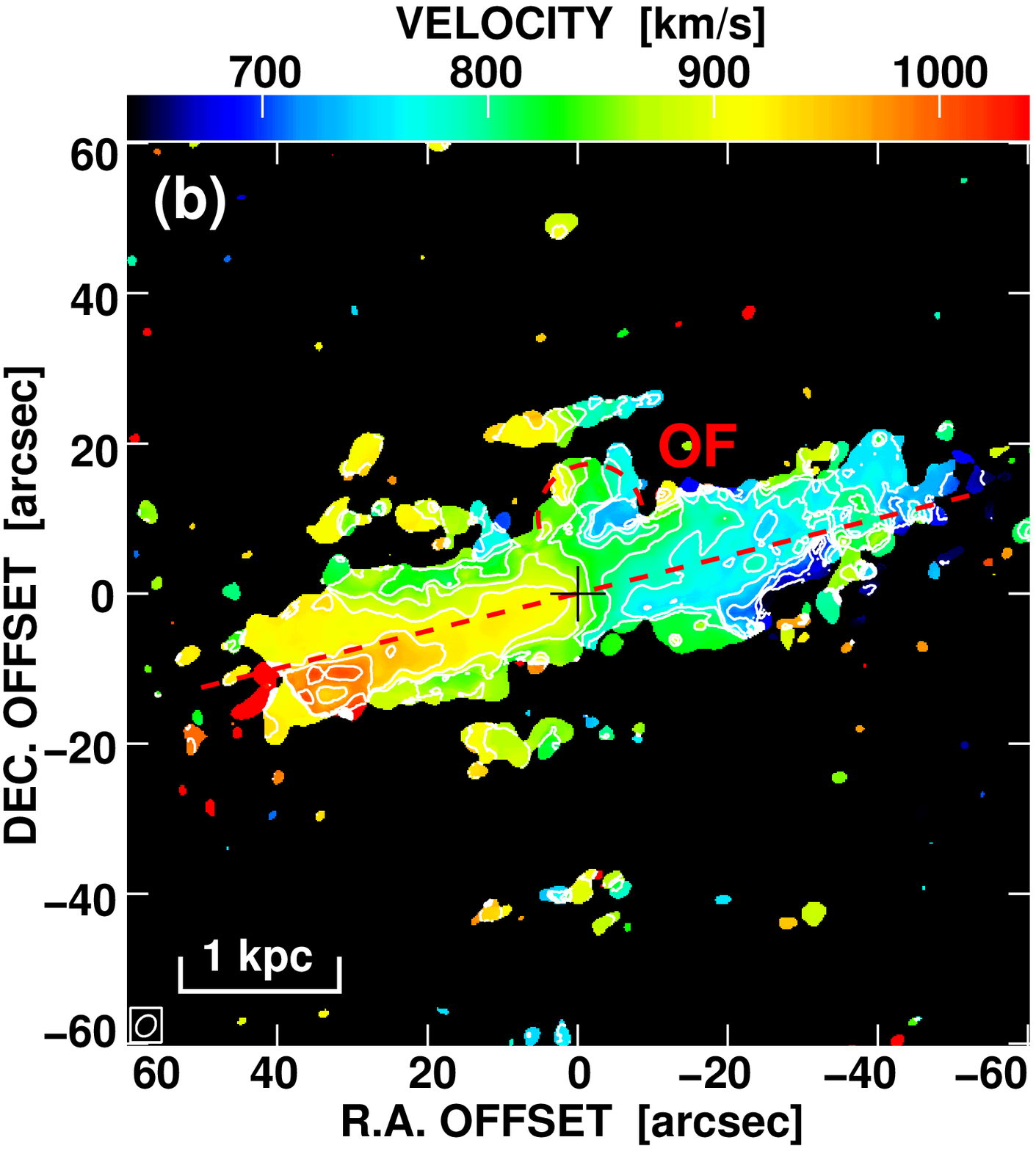}
\caption{
\label{mnt}
The NMA CO(1-0) integrated intensity (moment zero) and intensity weighted mean
velocity field (moment one) maps of NGC 3628.
The cross is the phase tracking center
and the synthesized beam is shown in the bottom-left corner of each map.
The central position and the synthesized beam size
 are the same as in Figure~\ref{chan}.
The red dashed line indicates the location of the galactic disk,
and the red curve indicates the molecular outflow {\it OF}.
(a) Moment 0 map.
The contour levels are 1, 3, 5, 10, 20, 50, 100, 150, and 180$\sigma$,
where 1$\sigma$ is 648.3~mJy~beam$^{-1}$~km~s$^{-1}$
    (= 9.05~K~km~s$^{-1}$).
(b) Moment 1 map.
The contour levels are from 640, 660, 680, $\dots$, and 1,040~km~s$^{-1}$,
increasing with 20~km~s$^{-1}$.
}
\end{figure*}

Figure~\ref{mnt}a is our NMA CO(1-0) integrated intensity map (moment zero map). 
Most of the CO emission is distributed along the galactic disk
 with a  position angle of 104$\arcdeg$.
The distribution of molecular gas in the galactic disk is not symmetric
with respect to the major axis.
The amount of 
the molecular gas above the major axis is more than that below the major axis.
Besides, both edges of the molecular disk are distorted.
The eastern edge of the disk shows a boxy-shape structure,
and the western edge of the disk is fragmented.
Both edges looks similar to the edges of the optical image
(Figure~\ref{opt-dss})
although the scale is different.
An extended emission located from the center with an offset of ($\Delta$R.A., $\Delta$Dec.) 
$\sim$ ($0\arcsec$, $+15\arcsec$)
labeled as {\it OF} is the real emission
because they also appear in the same position by mapping with uniform weighting.
However, those centers with position offsets of
$\sim$ ($0\arcsec$, $+25\arcsec$) and $\sim$ ($+7\arcsec$, $-20\arcsec$)
are considered as sidelobes
because the positions of these features
are either disappeared or shifted along with the beam patterns
by mapping with uniform weighting 
or by choosing different UV-ranges.
The existence of the strong sidelobes is due to 
 the UV-coverage of a low-declination source,
 which generates strong sidelobes along the north-south direction.

\subsubsection{Velocity Features}
\label{sec-pv}
\begin{figure}
\centering
\epsscale{1}
\plotone{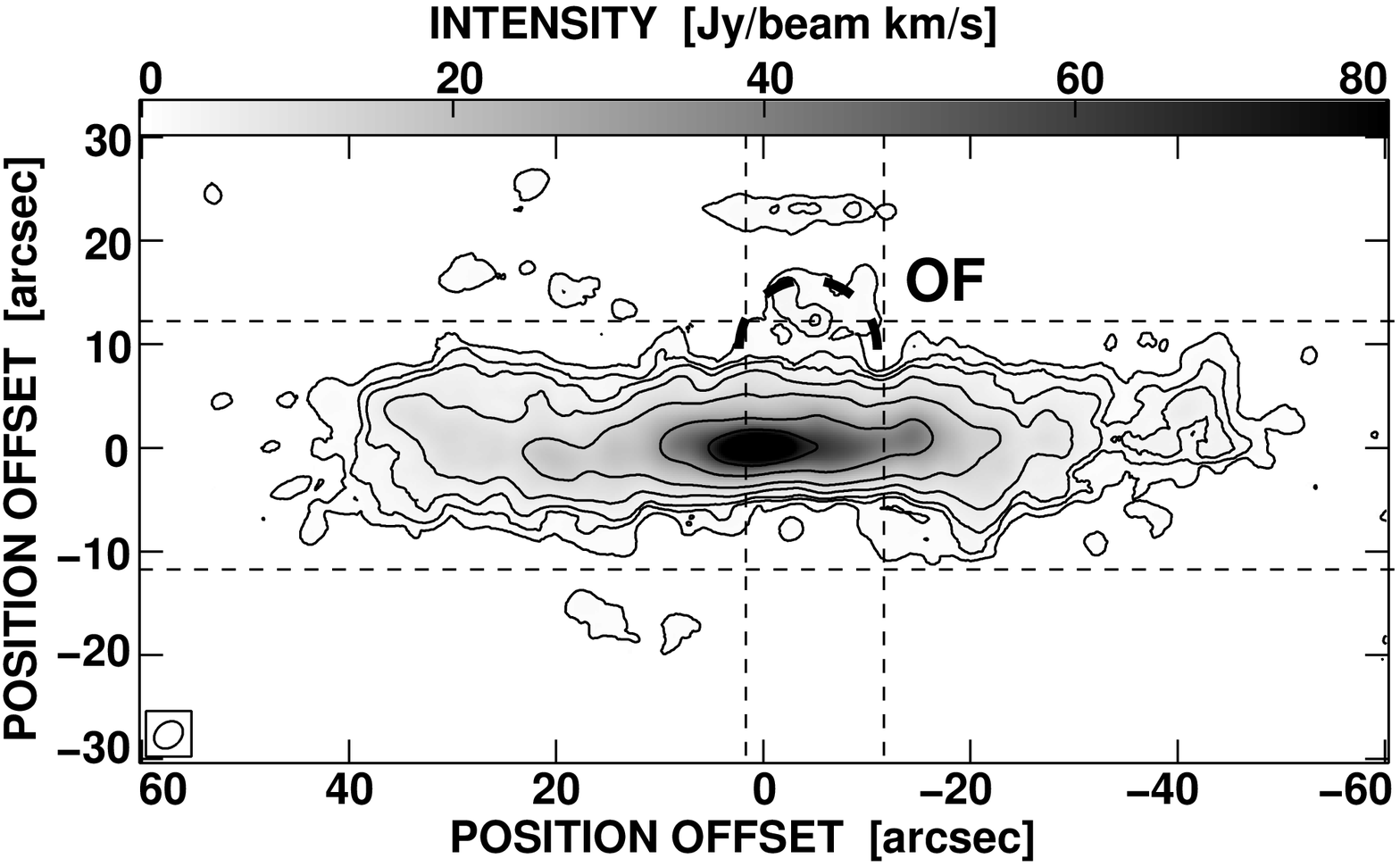}
\caption{
\label{m0rot14}
NMA CO(1-0) moment zero map after clockwise rotation of 14$\arcdeg$.
The contour levels are the same as in figure~\ref{mnt}a.
The two horizontal dashed lines, offset by $+12\arcsec$ and $-12\arcsec$
from the major axis, specify the range of major-axis $p-v$ diagrams
shown in Figure~\ref{pvmajors}.
The two vertical dashed lines, offset by $+1\arcsec$ and $-11\arcsec$
from the minor axis specify the range of {\it OF}.
Its minor-axis $p-v$ diagram is shown in Figure~\ref{pvminor}.
}
\end{figure}

Figure~\ref{mnt}b is the intensity weighted mean velocity map (moment one map).
From this map, we can see that
 there is a rigid-body rotation in the galactic center
 and a flat rotation in larger radii in the eastern part of the disk.
The velocity in the northwestern part of the disk seems to be a flat rotation,
 but the southwestern part of the disk does not show a clear flat rotation feature.
Besides, the velocity in the western part of the disk is twisted and entangled.
In order to further discuss the detail of these features,
   we first rotate the moment zero map until 
   the major axis of the galactic disk is the same with
   the horizontal direction,
   namely rotating clockwise for 14$\arcdeg$, as shown in Figure~\ref{m0rot14},
   and then we check the position-velocity (hereafter $p-v$) diagrams
   parallel or perpendicular to the major axis.

\begin{figure}
\centering
\epsscale{1}
\plotone{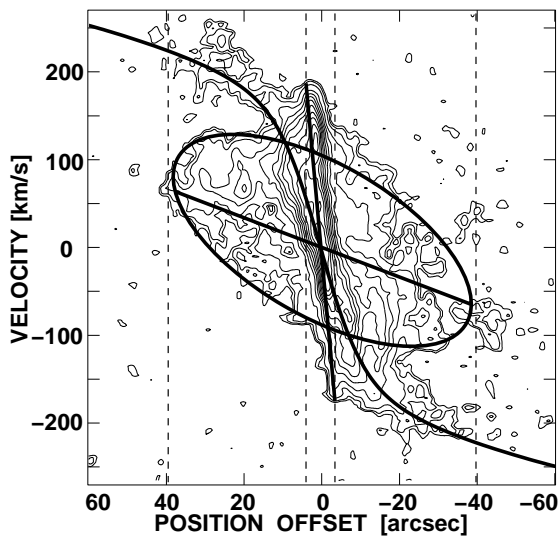}
\caption{
The $p-v$ diagram along the major axis
by averaging the data
between the positions $-1\arcsec$ and $+1\arcsec$ from the major axis.
The zero velocity corresponds to the systemic velocity of 840~km~s$^{-1}$.
The contour levels are
2, 3, 5, 10, 15, 20, ..., and 50$\sigma$,
where 1$\sigma$ is 9.09~mJy~beam$^{-1}$.
The solid curve between  $+60\arcsec$ and $-60\arcsec$
 indicates the rotation curve of the galactic disk.
The solid line between two dashed vertical lines at $+4\arcsec$ and $-4\arcsec$
 indicates the steep velocity gradient of the inner molecular disk.
The solid line between two dashed vertical lines at $+40\arcsec$ and $-40\arcsec$
 indicates the low velocity gradient of the outer molecular disk.
The solid ellipse indicates the velocity feature of the molecular gas bar.
\label{pvmajor0}
}
\end{figure}

Figure~\ref{pvmajor0} shows the major-axis $p-v$ diagram
 by averaging the $p-v$ diagrams within 2$\arcsec$
 (between $+1\arcsec$ and $-1\arcsec$ offsets) from the major-axis.
From Figure~\ref{pvmajor0}, we found that the galactic disk consists of several structures:
The curve between the position $+60\arcsec$ and $-60\arcsec$
indicates a rotation curve along the major axis of the galactic disk,
 namely, a rigid-body rotation between the positions $+10\arcsec$ and $-10\arcsec$,
 with a flat rotation beyond these positions;
 this suggests the existence of a large-scale molecular gas disk.
The solid line between the positions $+4\arcsec$ and $-4\arcsec$
 indicates a rigid-body rotation, suggesting the existence of an inner molecular disk.
The ellipse between the positions $+40\arcsec$ and $-40\arcsec$
 indicates the velocity feature more or less symmetric.
This feature is similar to the $p-v$ diagram of our Galaxy \citep{eng99},
 suggesting the existence of a molecular bar \citep{bin91,gar95} in this galaxy.
The solid line between the position $+40\arcsec$ and $-40\arcsec$ 
 will be discussed in the following paragraph.

\begin{figure*}
\centering
\epsscale{1}
\plotone{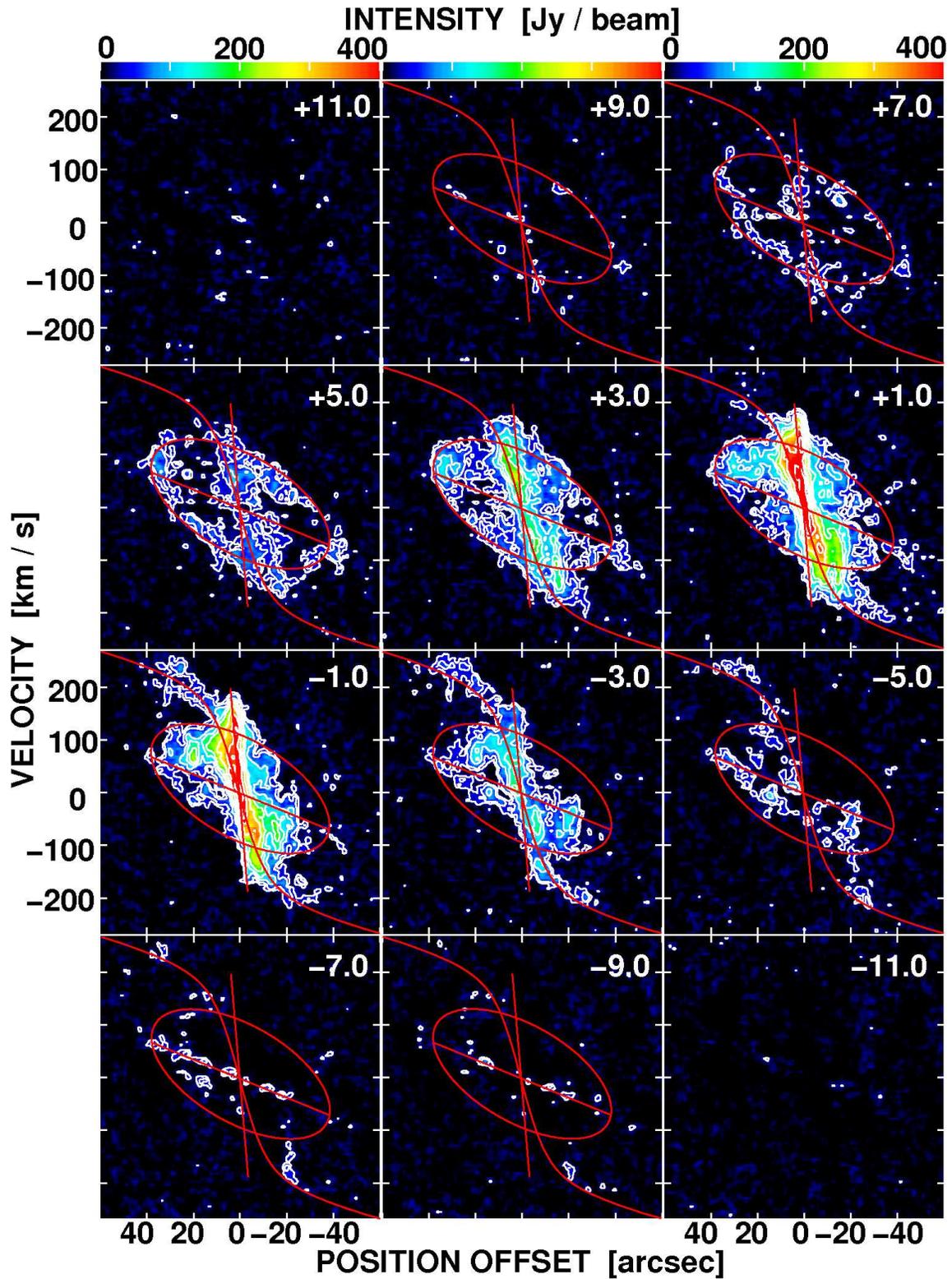}
\caption{
\label{pvmajors}
Averaged $p-v$ diagrams parallel to the major axis
in the range between the position offset of $-12\arcsec$ and $+12\arcsec$,
which are marked as two dashed horizontal lines in figure~\ref{m0rot14}.
Each $p-v$ diagram is averaged every 2$\arcsec$ with an interval of
2$\arcsec$.
The number in each $p-v$ diagram indicates the offset
from major-axis in unit of arcsec.
The solid lines, curves, and ellipses are the same as those shown in Figure~\ref{pvmajor0}.
The zero velocity corresponds to the systemic velocity of 840~km~s$^{-1}$.
The contour levels are  3, 5, 7, 10, 20, 30, 40, and 50$\sigma$,
where 1$\sigma$ is 9.09~mJy~beam$^{-1}$.
}
\end{figure*}

Figure~\ref{pvmajors} shows averaged $p-v$ diagrams parallel to the major axis with various offsets.
These $p-v$ diagrams are obtained by averaging 
 every 2$\arcsec$ in the range
 between the position offsets $+12\arcsec$ and $-12\arcsec$
 from the major axis (see Figure~\ref{m0rot14} for this range).
The red solid curve, lines, and ellipse are the same as those in Figure~\ref{pvmajor0}.
We found that the structure of the molecular disk is not symmetric
 in the north-south direction.
The inner disk can only be clearly seen in the position
 at the range between the offset $+1\arcsec$ and $-1\arcsec$.
The velocity feature of the bar structure can be seen in the position
 at the range between the offset $+7\arcsec$ and $-5\arcsec$.
In addition to these velocity features,
 we also found a rigid-rotation feature in the position at the range between the offsets
$-7\arcsec$ and $-9\arcsec$.
The emission is weak while the slope is shallow,
 and the features extend to $\pm40\arcsec$,
 so that this structure may be a rotating disk
 which is located at a large radius.
Here we call it an outer molecular disk.

\begin{figure}
\centering
\epsscale{1}
\plotone{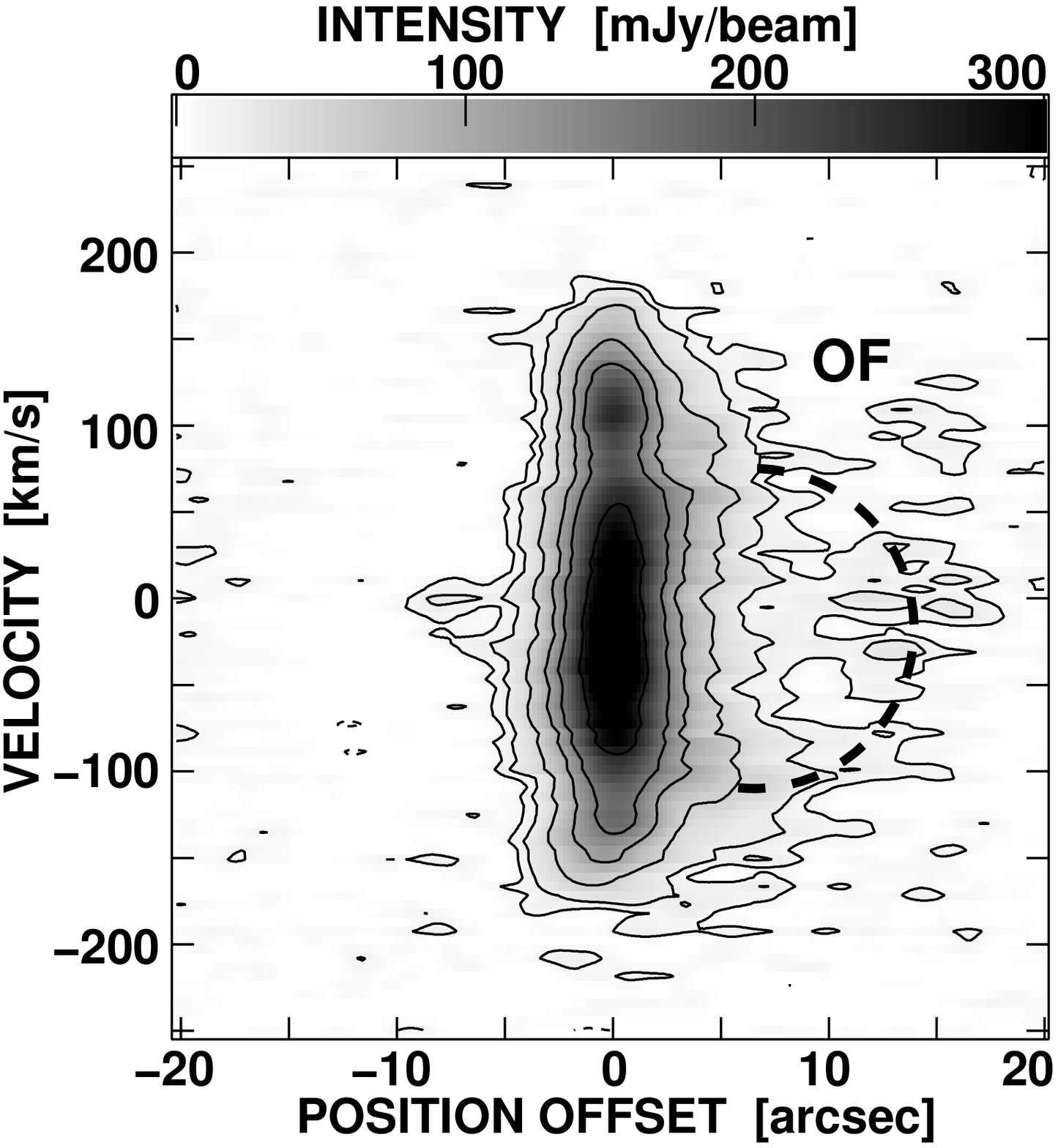}
\caption{
\label{pvminor}
The $p-v$ diagram along the minor axis averaged over
the range between the position offsets $+1\arcsec$ and $-11\arcsec$ from the minor axis,
which is between two dotted vertical lines in figure~\ref{m0rot14}.
The zero velocity corresponds to the systemic velocity of 840~km~s$^{-1}$.
The contour levels are 2, 4, 10, 20, 30, 50, and 70$\sigma$,
where 1$\sigma$ is 5.23~mJy~beam$^{-1}$.
}
\end{figure}

In order to see the velocity feature of the extended structure {\it OF}
 in the north of the galactic disk,
we averaged the $p-v$ diagrams along the minor axis
 in the range between the position offsets
 $+1\arcsec$ and $-11\arcsec$ on the major axis 
 (see Figure~\ref{m0rot14} for this range),
 and the averaged $p-v$ diagram is shown in Figure~\ref{pvminor}.
The strong emission is distributed along the position offset of $\sim$ 0$\arcsec$,
 indicating the location of the galactic disk. 
Northern part of the galactic disk (positive side of the position offset)
obviously shows an extended diffuse emission,
which corresponds to the extended structure {\it OF} in Figure~\ref{mnt}a and Figure~\ref{m0rot14}.
The velocity range of this diffuse emission
 is between $\sim-110$ -- +70 km~s$^{-1}$.
On the other hand,
 southern part of the disk (negative side of the position offset)
 shows no extended emission.

\subsubsection{Mass}
The total molecular gas mass can be estimated from the CO flux.
The total CO flux, $S_{\rm CO(1-0)}$,
measured from our moment zero intensity map (Figure~\ref{mnt}a) is 
$\sim$ 8.01 $\times$ 10$^3$~Jy~km~s$^{-1}$ (1.12~$\times~10^5$~K~km~s$^{-1}$~pc$^{2}$
 from the central $\approx80\arcsec~\times~20\arcsec$ of the galaxy).
The total flux is four times larger than that measured by \citet{irw96} 
($\sim$ 2.05 $\times$ 10$^3$ Jy~km~s$^{-1}$)
 from the same region of the galaxy.
 This is because 
 our high sensitivity NMA observation 
 provides a better sensitivity than \citet{irw96}
(as mentioned in Sec.~\ref{n3628-overall-prop})
so that we can detect much  weaker emission.
The total molecular gas mass derived from our CO data,
 in units of M$_\sun$, can be calculated as
\begin{eqnarray}
\label{massh2}
M_{\rm H_2} &=& 1.2 \times 10^4 \times D^2 \nonumber \\
 && \times\ S_{\rm CO(1-0)}\times \frac{X_{\rm CO}}{3.0 \times 10^{20}}, \\
\label{massmol}
M_{\rm gas} &=& 1.36 \times M_{\rm H_2},
\end{eqnarray}
where $D$ is the distance in units of Mpc,
$X_{\rm CO}$ is the CO-to-H$_2$ conversion factor in units of cm$^{-2}$~K~(km~s$^{-1})^{-1}$,
and the factor 1.36 is a coefficient convert
from molecular hydrogen gas mass to total gas mass, including helium \citep{sak95}.
Here, we used the value of $X_{\rm CO}$ = 1.4~$\times$~10$^{20}$~cm$^{-2}$~K~km~s$^{-1}$ \citep{mat00},
 assuming the condition of the molecular gas is similar to that of M82.
Thus the total gas mass, $M_{\rm gas}$,
is $\sim$ 1.8 $\times$ 10$^9$~M$_{\sun}$.

The dynamical mass, $M_{\rm dyn}$ within the radius of $r$,
can be derived from the rotational curve.
Figure~\ref{pvmajor0} shows that the maximum rotational velocity, 
$v_{rot}\sim~220$~km~s$^{-1}$,
appears at the location of radius $\sim$~40$\arcsec$ (1.5~kpc).
We assume a circular rotation for the galactic disk,
$v_{rot}^2 = GM_{\rm dyn}/r$.
Therefore, the dynamical mass within 1.5~kpc radius can be estimated as
$\sim$ 2 $\times$ 10$^{10}$~M${_\sun}$.
This indicates that the gas mass to dynamical mass ratio of 10\%,
which is the typical value for the central regions of galaxies 
(c.f., from 4\% for Sa galaxies to 25\% for Scd galaxies; \citealt{you91}).

\subsection{CXO X-ray Archive Data}
\subsubsection{Image of the Diffuse Emission}
\label{sec-xray-img}
In order to compare the CXO X-ray data with the NMA CO(1-0) data in different scale,
 we made two types of smoothed images.

\begin{figure*}
\centering
\epsscale{1}
\plottwo{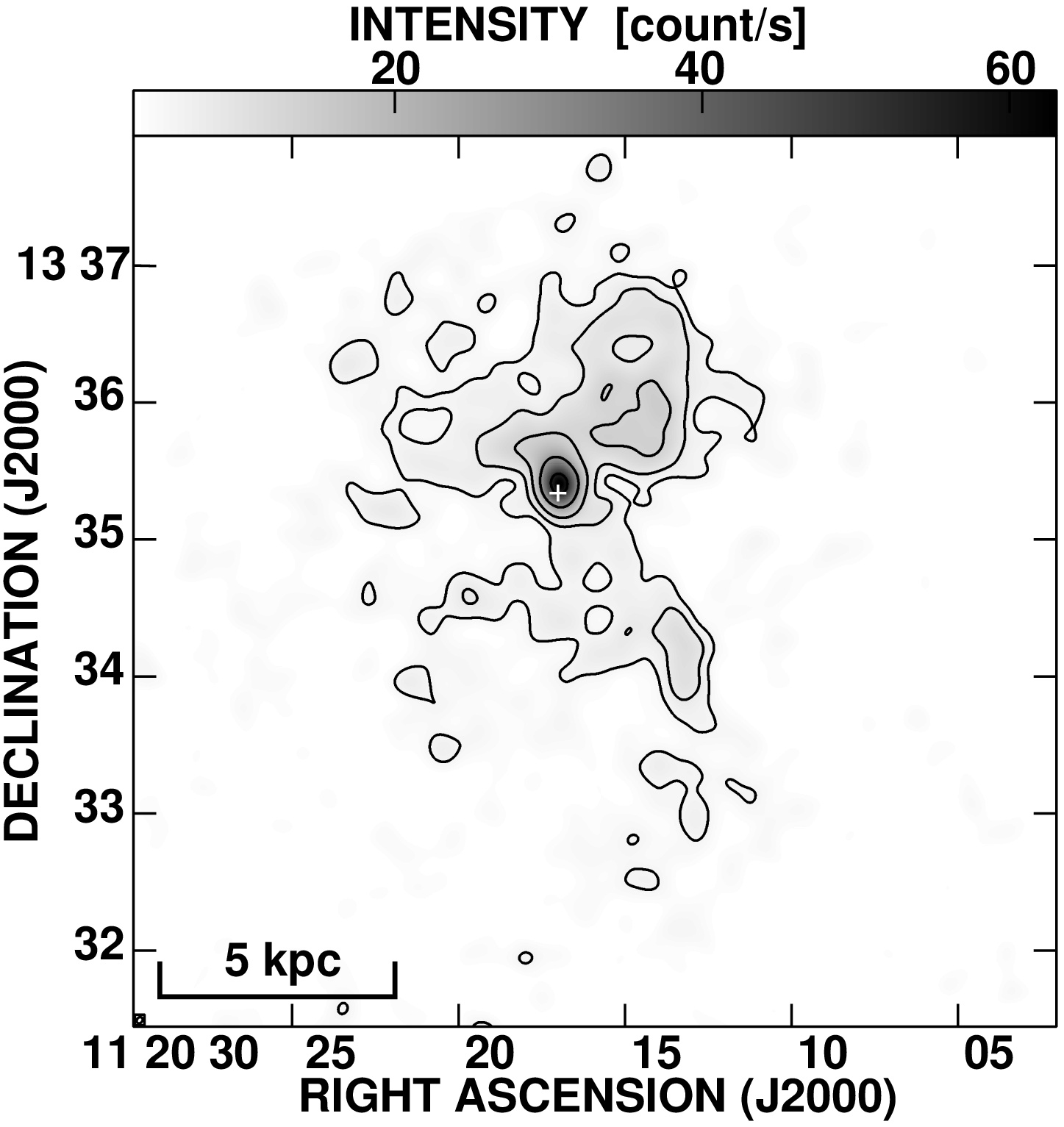}{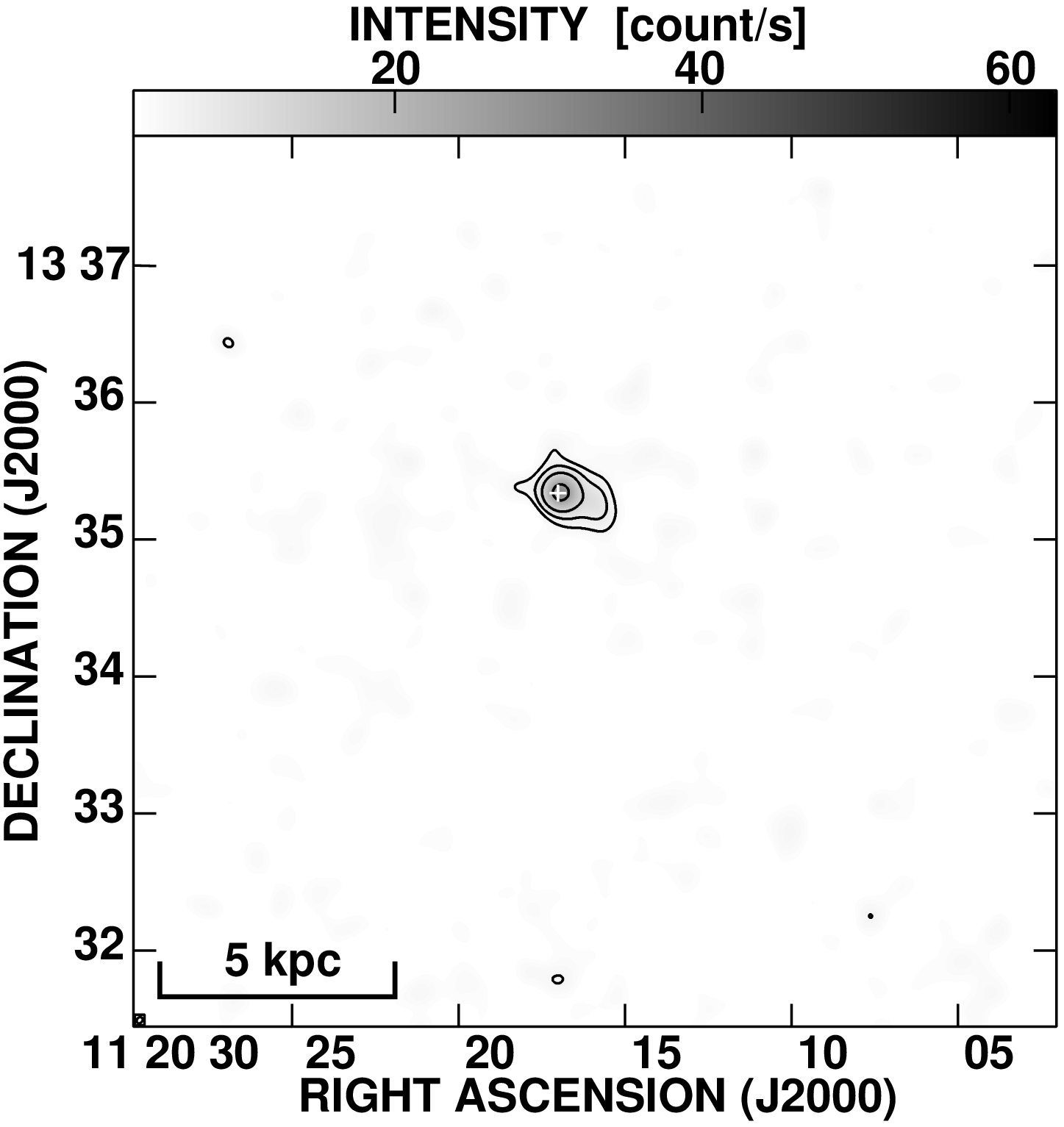}
\caption{
\label{xray-img}
Smoothed (by 25$\arcsec$) CXO X-ray image after removing point sources.
The cross is the phase tracking center of the NMA CO(1-0) data at
$\alpha{\rm (J2000)} = 11^{\rm h}20^{\rm m}04{\fs}863$ and
$\delta{\rm (J2000)} = 13\arcdeg30\arcmin50\farcs49$.
(a) Soft band (0.3 -- 2.0~keV).
The contour levels are 2, 3, 5, 10, and 20 $\times$ 2.6757 counts~s$^{-1}$.
(b) Hard band (2.0 -- 7.0~keV).
The contour levels are 2, 3, 5, and 10 $\times$ 2.5802 counts~s$^{-1}$.
}
\end{figure*}

\begin{figure*}
\centering
\epsscale{1}
\plottwo{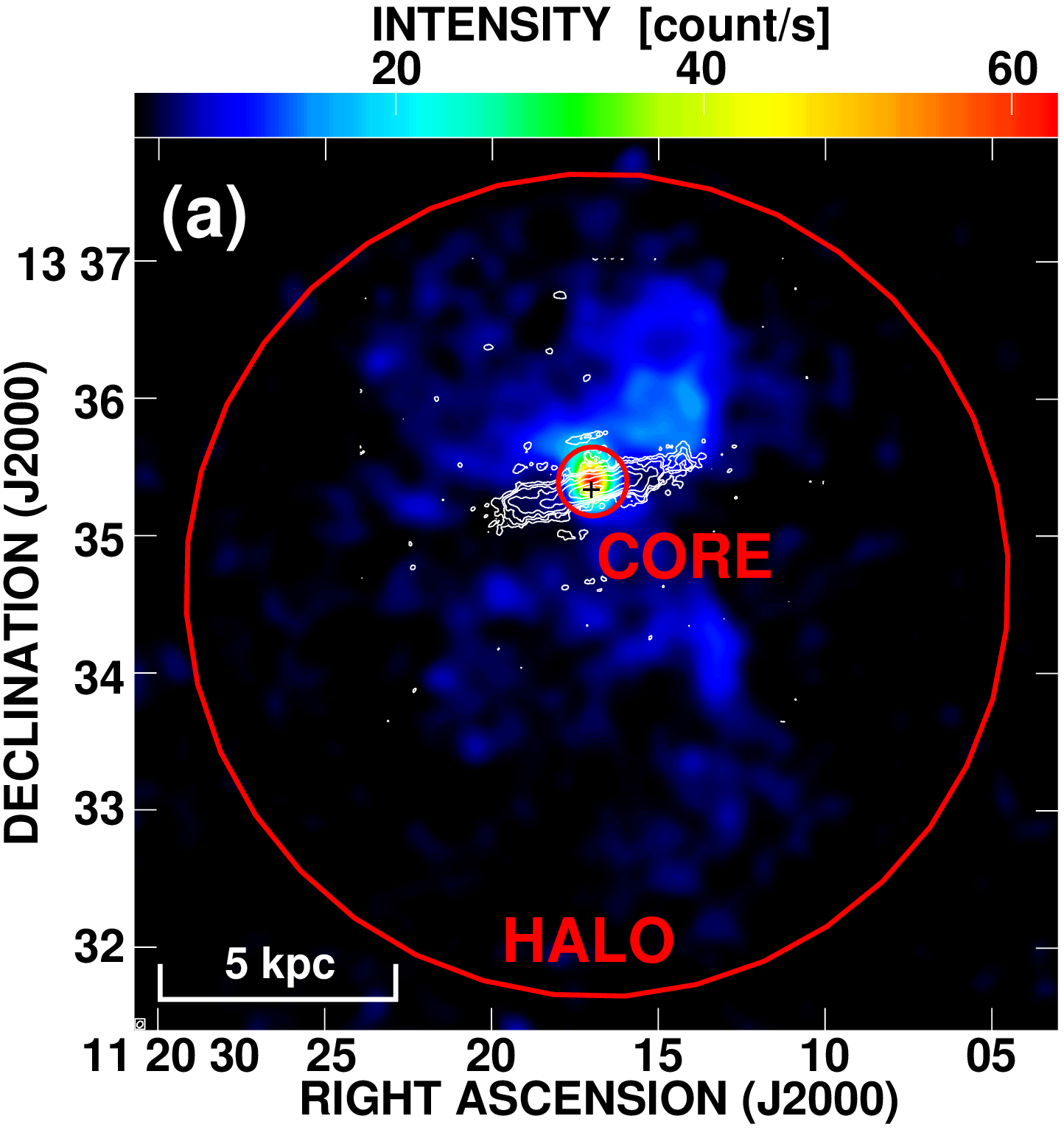}{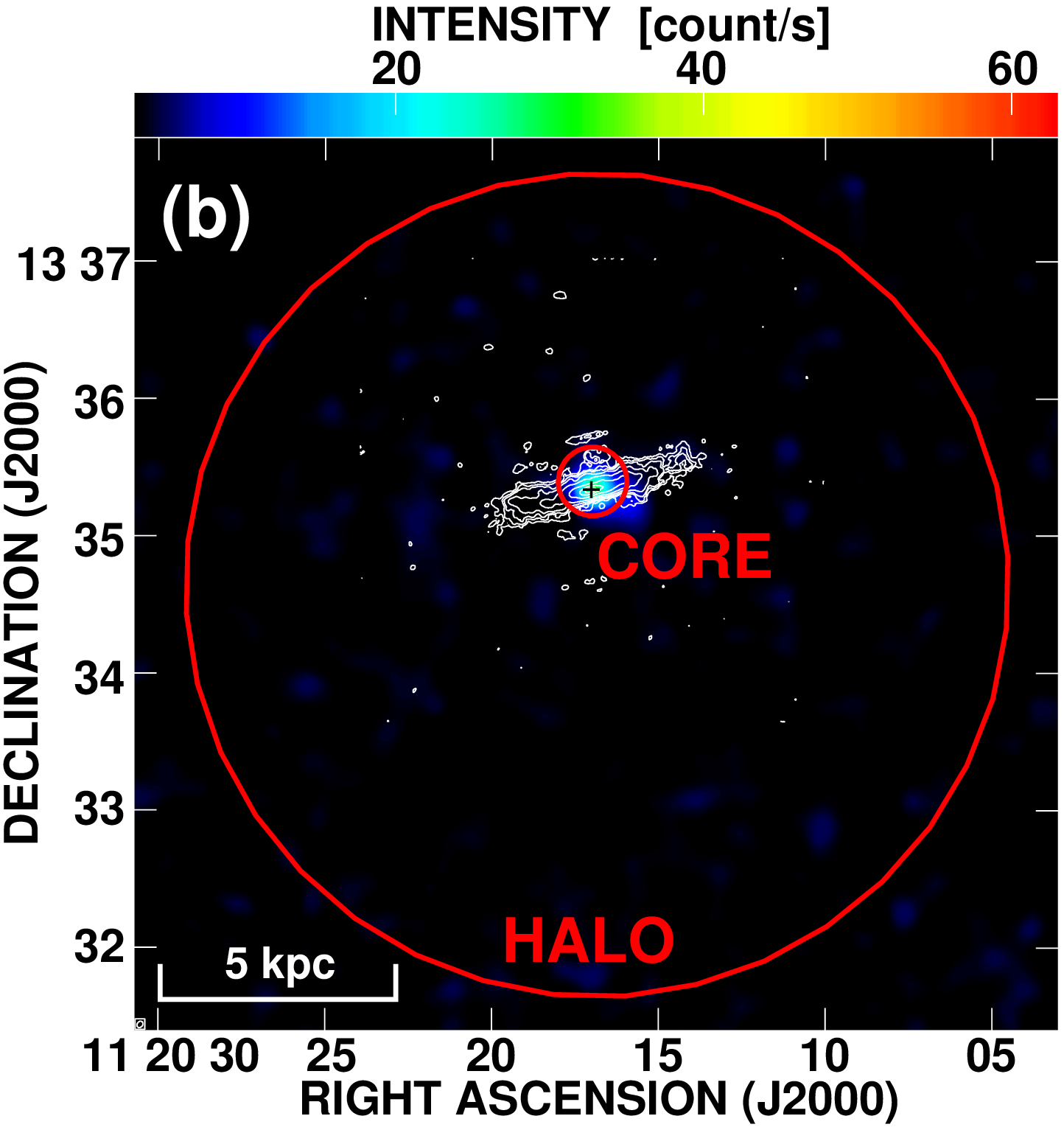}
\plottwo{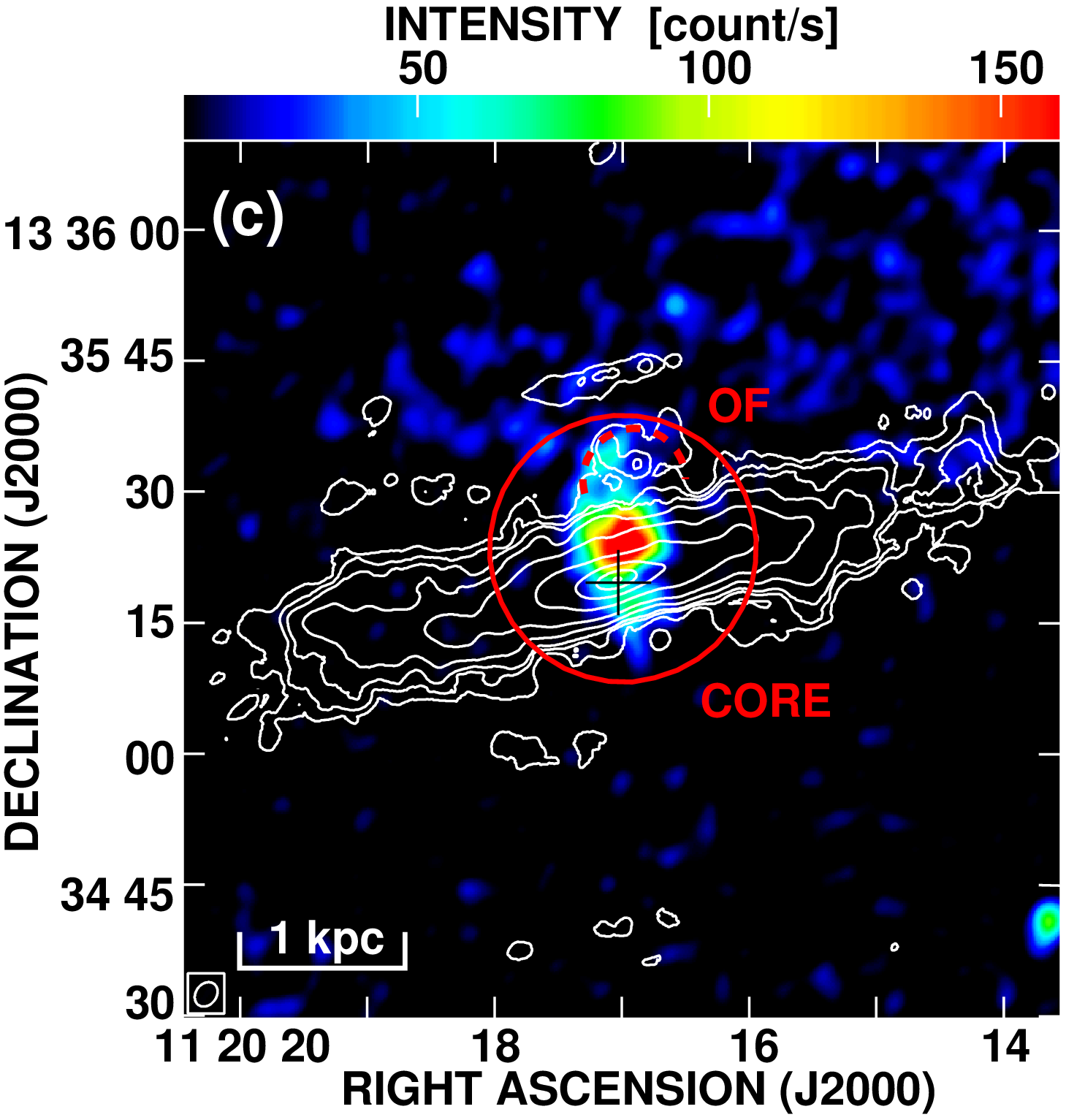}{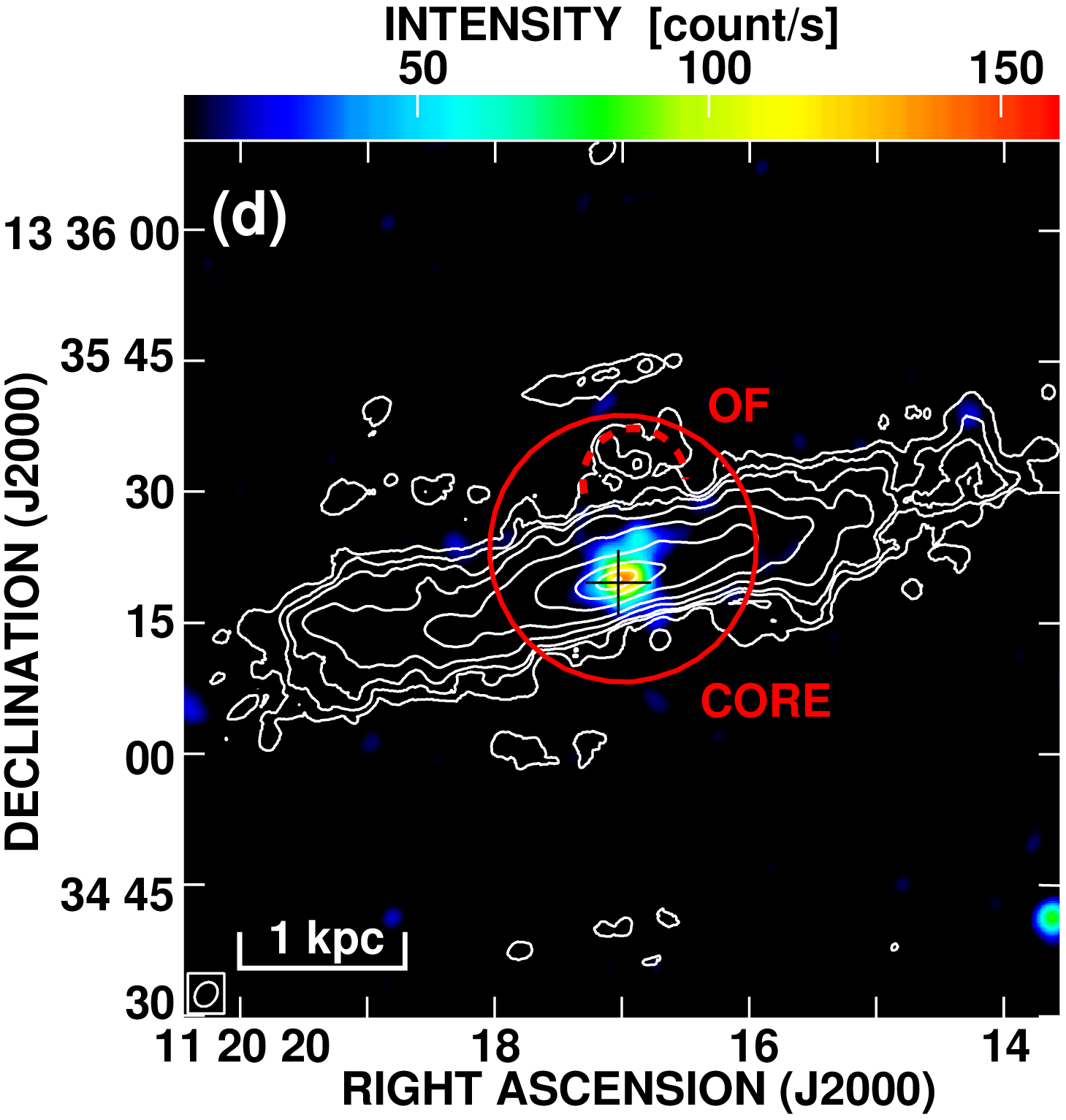}
\caption{
\label{xray-co}
The spatial correlation between
 the NMA CO(1-0) data (contours) and
 the CXO X-ray data (color scale).
The contour levels are the same as those in Figure~\ref{mnt}.
The cross is the phase tracking center of the NMA CO(1-0) data.
The large and small red solid circles are the regions {\it HALO} and {\it CORE}, respectively.
The red dashed curve is the {\it OF}.
(a) CO(1-0) contours overlaid on the soft band (0.3 -- 2.0~keV) X-ray image.
(b) CO(1-0) contours overlaid on the hard band (2.0 -- 7.0~keV) X-ray image.
(c) Zoom-in image of the central 2$\arcmin\times2\arcmin$ region of (a).
(d) Zoom-in image of the central 2$\arcmin\times2\arcmin$ region of (b).
}
\end{figure*}

To image the large scale diffuse structure,
 we used the CIAO tool ACONVOLVE to convolve the point-source-removed X-ray image with a Gaussian
that is smoothed by 25$\arcsec$
in 2 energy bands,
soft (0.3 -- 2.0~keV) and hard (2.0 -- 7.0~keV) bands.
We also tried to smooth with 20$\arcsec$,
 but the diffuse emission cannot be displayed clearly.
The smoothed soft and hard X-ray images are shown
 in Figure~\ref{xray-img}a and Figure~\ref{xray-img}b, respectively.
These two images are very similar to those published by \citet{str01},
suggesting that our imaging is consistent with theirs.
In soft-band X-ray emission (Figure~\ref{xray-img}a),
the strongest intensity is very close to the galactic center,
 but shifted a little bit toward the north.
An asymmetric plasma outflow
 located above and below the galactic center is clearly seen.
The northern plasma outflow has a stronger intensity and wider distribution than the southern outflow,
while the southern outflow extends longer distance than the northern one.
The diffuse emission spreads over $\sim$ 7 -- 10 kpc scale in both north and south of the galactic disk.
In hard-band X-ray emission (in Figure~\ref{xray-img}b),
 the strongest intensity is at the galactic center.
The emission is roughly concentrated in the galactic center,
 and we do not see clear evidence of the outflow feature.
The overlaid images between the CXO X-ray data and the NMA CO(1-0) data
 are shown in Figure~\ref{xray-co}a and Figure~\ref{xray-co}b.

To image the small scale diffuse structure, 
 we used AIPS task CONVL to convolve the X-ray data to the same beam size as
 ($3\farcs01 \times 2\farcs36$) as
 the NMA CO(1-0) data in 2 energy bands.
The high resolution soft and hard X-ray images are shown
 in Figure~\ref{xray-co}c and Figure~\ref{xray-co}d, respectively.
In the higher resolution soft-band X-ray emission map (Figure~\ref{xray-co}c),
 the distribution of the plasma outflow near the galactic center is roughly located
 in similar region as the extended molecular gas feature {\it OF},
 while the peak intensity of  the X-ray emission
 is located a little bit toward the north.
In the higher resolution hard-band X-ray emission map (Figure~\ref{xray-co}d),
 the distribution is concentrated in the galactic center.

\subsubsection{Spectral Analysis of the Diffuse Emission}
In order to compare the physical properties of plasma gas to that of molecular gas,
 we used XSPEC to fit the spectra of the X-ray data to
 obtain the temperature $kT$, and the normalization term $Norm$,
 which can be used to estimate 
 mean electron density, plasma mass, thermal pressure, and thermal energy
 (the details will be described in Sec.~\ref{sec-ion-prop}).

\begin{figure*}
\centering
\epsscale{1}
\plottwo{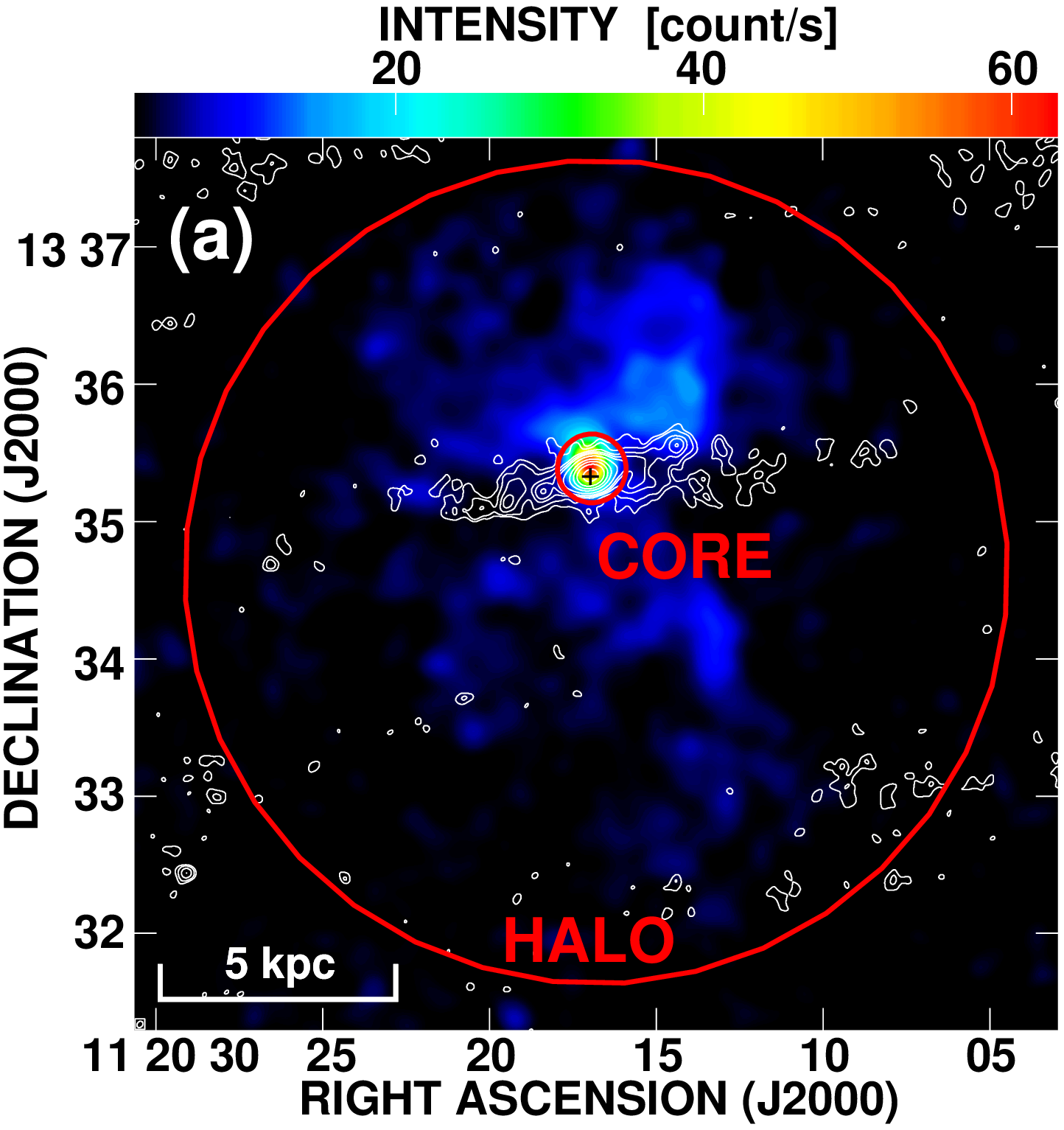}{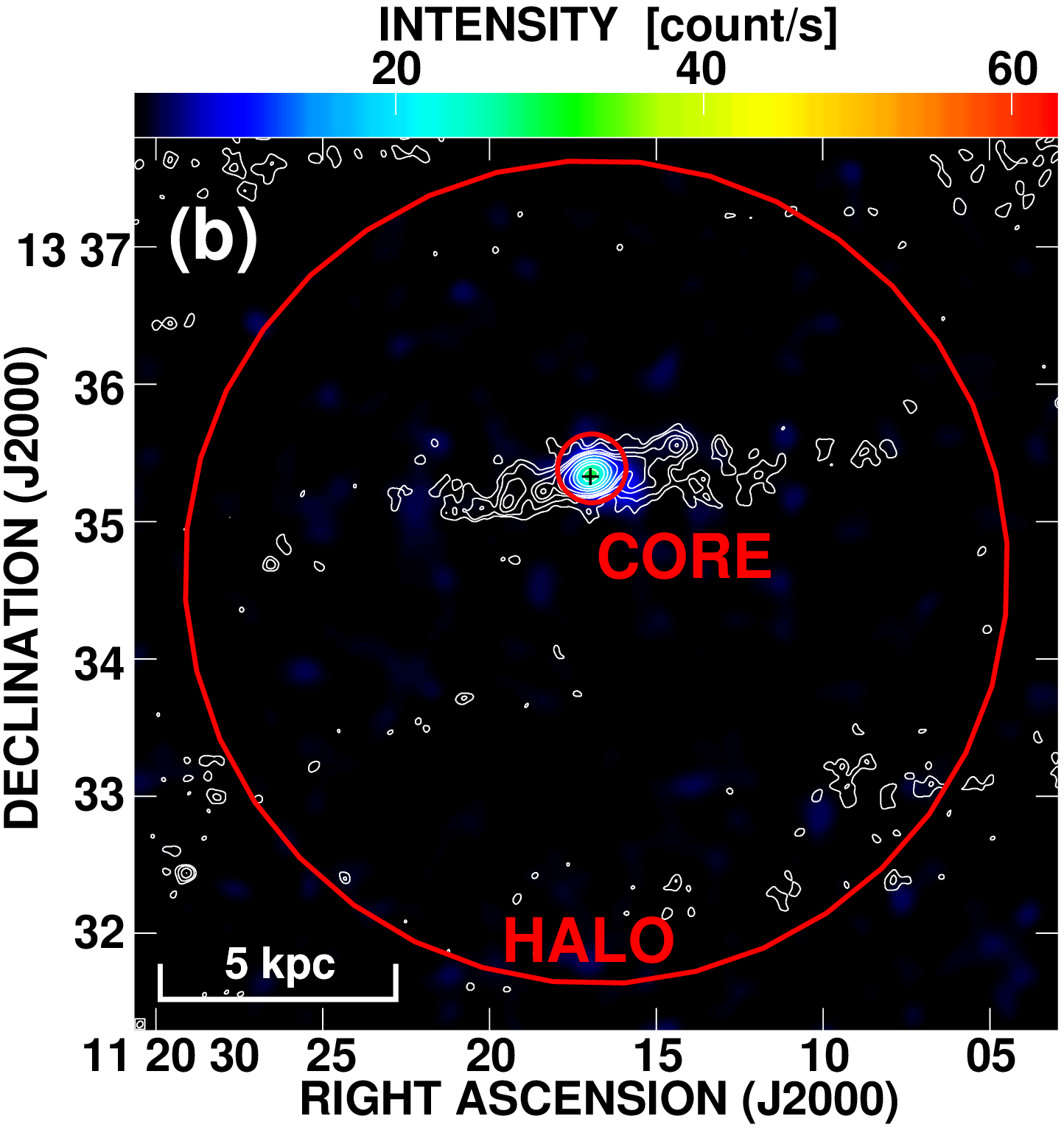}
\caption{
\label{xray-HI}
Smoothed (by 25$\arcsec$) CXO X-ray image (color) overlaid on the
H{\scriptsize I} VLA FIRST data (contour).
The cross is the phase tracking center of the NMA CO(1-0) data.
(a) H{\scriptsize I} contours overlaid on the soft-band (0.3 -- 2.0~keV) X-ray image.
(b) H{\scriptsize I} contours overlaid on the hard-band (2.0 -- 7.0~keV) X-ray image.
The contour levels are 2, 3, 5, 7, 10, 20, 50, 100, 200, and 500$\sigma$,
where 1$\sigma$ is  0.15 mJy~beam$^{-1}$.
}
\end{figure*}

For this spectral analysis, we first define spectral fitting regions.
Figure~\ref{xray-HI}a and Figure~\ref{xray-HI}b show 
 the VLA FIRST H{\small I} archive image \citep{bec95} 
 overlaid on the soft and hard X-ray image, respectively.
The H{\small I} line emission is distributed along the galactic disk.
After comparing the distribution of X-ray emission, H{\small I} emission,
 and the extended molecular gas {\it OF}, 
 we specified/defined two fitting regions, {\it CORE} and {\it HALO}.
For simplicity, we defined the region {\it CORE} as a circular area
 with a radius of $\sim$~0$\farcm$3 (0.67~kpc),
 which includes the central strong X-ray emission in the H{\small I} disk
 and the feature {\it OF} (Figure~\ref{xray-co}a, \ref{xray-co}c, and \ref{xray-HI}a).
Meanwhile, we defined the region {\it HALO} as a circular area
 with a radius of $\sim$~3$\farcm$0 (6.72~kpc), which 
 includes all diffuse X-ray emission but excludes the region {\it CORE}.

The temperature $kT$, and the normalization term $Norm$,
depend on the absorption column density
 and the metal abundance of the galaxy.
The absorption column density of atomic hydrogen, $N_{\rm H}$,
 can be fitted with the photoelectric absorption model, WABS.
In {\it CORE},
 the absorption is affected by both the host galaxy and the Milky Way.
Since the absorption in the host galaxy is unknown,
 we left $N_{\rm H}$ as a free parameter.
In {\it HALO}, 
 X-ray emission has no counterpart in the H{\small I} disk,
 thus we only consider the absorption from the Milky Way.
We therefore fixed $N_{\rm H}$ 
 as Milky Way absorption column density,
 0.02~$\times$~10$^{22}$~cm$^{-2}$ \citep{kal05,dic90}.

\begin{figure*}
\centering
\epsscale{1}
\plottwo{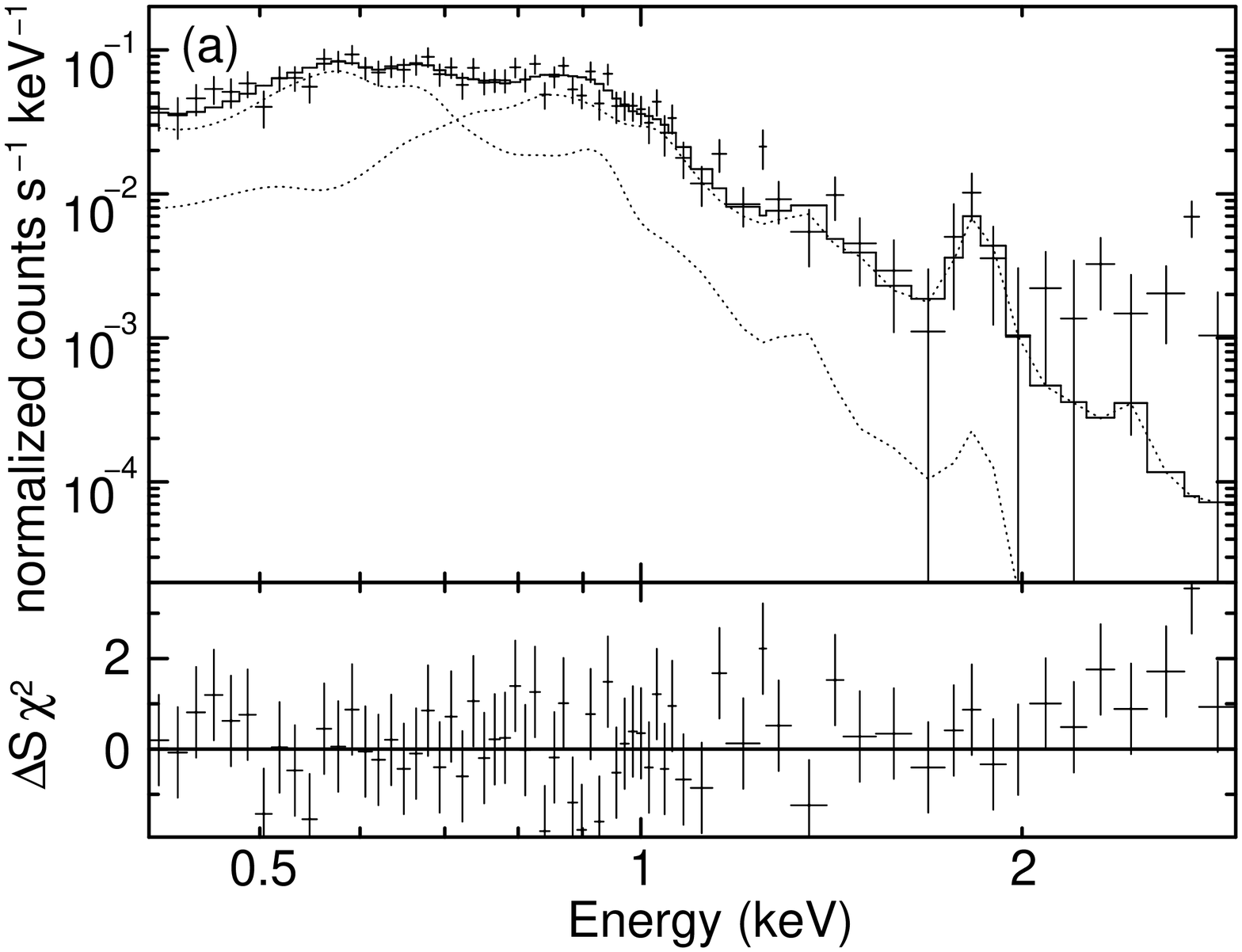}{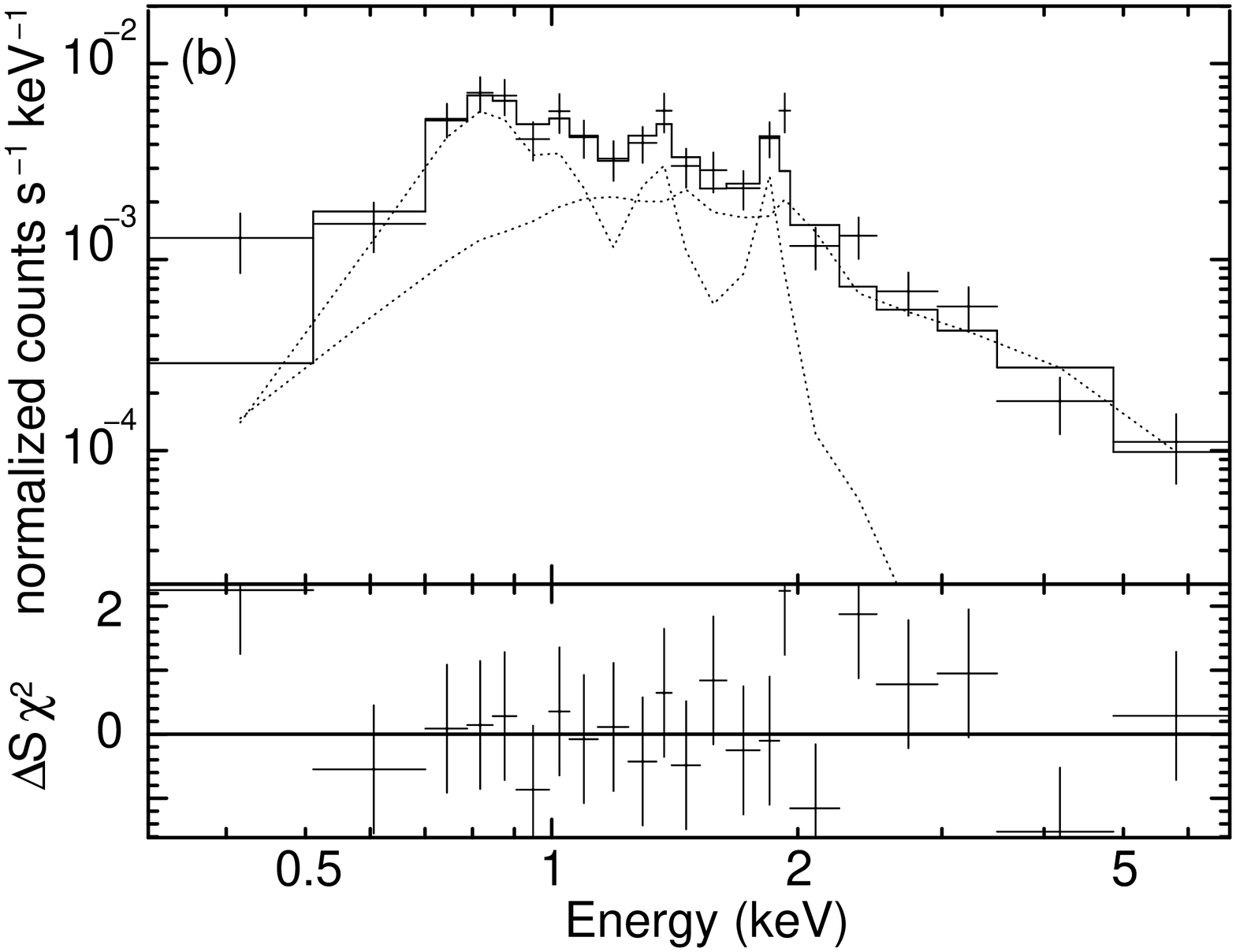}
\caption{
\label{xray-spec}
The CXO X-ray data with the fitted XSPEC model.
The top plots in each figure are the data (crosses) with the fitted model (dot for individual model and histogram for combined model).
The bottom plots in each figure are $\chi^2$.
(a) Spectrum extracted from the {\it HALO} region with
the fitting model of WABS*(VMEKAL + VMEKAL).
(b) Spectrum extracted from the {\it CORE} region with
the fitting model of WABS*(VMEKAL + VMEKAL).
The fitting parameters are shown in Table~\ref{xray-fit}.
}
\end{figure*}

For the metal abundance,
 we identify O ($\sim$ 0.5 -- 0.7~keV),
 Si ($\sim$ 1.8 -- 1.9~keV), and Fe ($\sim$ 0.7 -- 1.0~keV)
 in the {\it HALO} spectrum (Figure~\ref{xray-spec}a),
 and  Mg ($\sim$ 1.3 -- 1.4~keV), Si ($\sim$ 1.8 -- 1.9~keV), and Fe ($\sim$ 0.7 -- 1.0~keV)
 in the {\it CORE} spectrum (Figure~\ref{xray-spec}b).
The existence of these lines suggests that the emission originates from 
  optically-thin thermal plasma \citep{inu05}.
Thus we can choose a thermal-plasma model from XSPEC, 
i.e.,  VMEKAL or VAPEC.

Since the data quality of the {\it HALO} spectrum 
is poor at energy lower than 0.4~keV and higher than 1.3~keV,
we intended to exclude these data.
In addition to considering the metal abundance of Si at $\sim$~1.8 -- 1.9~keV,
 we include data at higher energy side till 3.0~keV.
Therefore, the soft-band energy range is 0.4 -- 3.0~keV.
We begin to fit the spectrum with an absorbed single-temperature thermal-plasma model.
However, the spectrum cannot be fitted well neither by WABS*VMEKAL nor by WABS*VAPEC model.
Thus we tried to fit the spectrum via an absorbed two-temperature thermal-plasma model,
 i.e., WABS*(VMEKAL + VMEKAL) or WABS*(VAPEC + VAPEC).
We also fix the metal abundances of the higher temperature component to those
 of the lower temperature component.
The fitting results show that 
 the model WABS*(VMEKAL + VMEKAL) fits better than
the model WABS*(VAPEC + VAPEC). 
The fitting parameters in the model WABS*(VMEKAL + VMEKAL) 
for both {\it HALO} and {\it CORE} regions are listed in Table~\ref{xray-fit}.

In {\it HALO}, the metal abundances fitted by the  model 
 are $\sim$~0.5~$Z_\sun$ for O, $\sim$~2.8~$Z_\sun$ for Si, 
 and $\sim$~0.4~$Z_\sun$ for Fe, 
 where $Z_\sun$ is the solar abundance.
Two temperatures are $\sim$~0.21~keV and 0.60~keV.
In {\it CORE},
the metal abundances fitted by the model 
 are $\sim$~5.0~$Z_\sun$ for Mg,
  $\sim$~13.5~$Z_\sun$ for Si,
 and $\sim$~0.8~$Z_\sun$ for Fe.  
Two temperatures are $\sim$~0.47~keV and 7.3~keV.
We notice that both the metal abundance of Mg and Si 
are extremely high.
These values are likely not physical.
Besides, their large uncertainties
 (1.5 -- 25.9~$Z_\sun$ for Mg and 5.3 -- 50.8~$Z_\sun$ for Si)
suggest that the fitted values are not accurate enough.
One possible reason is that the CCD material contaminates
 the detection of Si.
Another possible reason is that the photon counts, $\sim 1000$ counts,
 are too low to get a reasonable fitting results for Mg and Si.
We fix the metal abundances of Mg and Si as unity
and re-fit the spectrum again.
This does not change $kT$ and {\it Norm} too much. 
Therefore, we could ignored the influence from high abundances of Mg and Si.

\subsubsection{The Properties of Plasma Gas}
\label{sec-ion-prop}
From  $kT$ and $Norm$, we could derive 
mean electron density $n_{\rm e}$, plasma mass $M_{\rm plm}$, 
thermal pressure $P_{\rm plm,thm}$, and thermal energy $E_{\rm plm,thm}$.
The definition of $Norm$ is 
\footnote[1]{The definition is from XSPEC website,\\
   http://heasarc.gsfc.nasa.gov/docs/xanadu/xspec/manual/ }
\begin{eqnarray}
\label{eq-norm}
Norm &=& \frac{10^{-14}}{4\pi(D_{\rm A}(1+z))^2}\ EI, 
\end{eqnarray}
where $EI$ is the emission integral,
\begin{eqnarray}
EI &=& \int n_{\rm e}n_{\rm H}fdV, 
\end{eqnarray}
 $D_{\rm A}$ is angular diameter distance,
$z$ is redshift, 
$V$ is the volume of plasma, 
$n_{\rm H}$ is the hydrogen density,
and $f$ is the volume filling factor.
For nearby galaxies, $D_{\rm A}$ is simply the distance $D$.
Assuming that the electron and hydrogen have the same density,
i.e., $EI = \int n_{\rm e}^2fdV$,
we can derive the following plasma properties. 

The mean electron number density of plasma is,
\begin{eqnarray}
\label{eq-ei-n}
n_{\rm e} &=& (\frac{EI}{Vf})^{1/2}.
\end{eqnarray}
The mass of plasma is,
\begin{eqnarray}
\label{eq-ei-m}
M_{\rm plm} &=& \rho_{\rm plm} V = m_{\rm p}n_{\rm e} V,
\end{eqnarray}
where $\rho_{\rm plm}$ is the mass density of plasma,
and $m_{\rm p}$ is the proton mass.
The thermal pressure of plasma is,
\begin{eqnarray}
\label{eq-ei-p}
P_{\rm plm,thm} &=& \frac{\rho_{\rm plm}}{\mu m_{\rm p}}\ kT= 2n_{\rm e}kT,
\end{eqnarray}
where $\mu$ is the mean molecular weight.
For a fully ionized hydrogen,
 there are two particles for every proton and $\mu = 1/2$.
The thermal energy of plasma is,
\begin{eqnarray}
\label{eq-ei-e}
E_{\rm plm,thm} &=& \frac{3}{2}\ P_{\rm plm,thm}V = 3n_{\rm e}VkT
\end{eqnarray}
 \citep{car96}.

We applied $z$ = 0.0028 base on the systemic velocity measured
from our NMA CO(1-0) data.
The derived plasma parameters from 
the two-temperature model
in {\it HALO} and {\it CORE} regions
are shown in Table~\ref{xray-prop}.

The electron densities of the lower and the higher temperature component in {\it CORE}
 are 9.5$f^{-1/2}$ $\times$ 10$^{-3}$~cm$^{-3}$
 and 2.0$f^{-1/2}$ $\times$ 10$^{-2}$~cm$^{-3}$, respectively,
 about one order of magnitude higher than those in {\it HALO}.
The plasma masses of the lower and the higher temperature component in {\it CORE} 
 are 3.0$f^{1/2}$ $\times$ 10$^{5}$~M$_{\sun}$
 and 6.4$f^{1/2}$ $\times$ 10$^{5}$~M$_{\sun}$, respectively,
 about one to two orders of magnitude smaller than those in {\it HALO}.
The plasma thermal pressures of the lower and the higher temperature component in {\it CORE}
 are 1.4$f^{-1/2}$ $\times$ 10$^{-11}$~dyne~cm$^{-2}$
 and 4.8$f^{-1/2}$ $\times$ 10$^{-10}$~dyne~cm$^{-2}$, respectively,
 about one to two orders of magnitude higher than those in {\it HALO}.
The plasma thermal energies of the lower and the higher temperature component in {\it CORE}
 are 0.8$f^{1/2}$ $\times$ 10$^{54}$~erg
 and 2.7$f^{1/2}$ $\times$ 10$^{55}$~erg, respectively,
 about one order of magnitude smaller than those in {\it HALO}.

\subsection{The Extended Structure above Galactic Disk}

\label{n3628-mol-prop}
Our NMA CO(1-0) observations detected an extended structure {\it OF} that
 appears only in the north of the galactic disk for the first time (Figure~\ref{mnt}a).
After comparing
 the morphology between the NMA CO(1-0) data and the CXO X-ray data
(Figure~\ref{xray-img}a,
 Figure~\ref{xray-co}c, and Figure~\ref{xray-co}d),
 we found that the distribution of {\it OF} is roughly matching
 the northern plasma outflow near the galactic center ({\it CORE}),
 and seems to be the ejection point of the northern large-scale plasma outflow.
The intensity peak of X-ray emission is a little bit
 toward the north of {\it OF},
 which is similar to 
NGC~2146 \citep{alt09}.
Combining the kinematic feature of {\it OF} 
 in the $p-v$ diagram (Figure~\ref{pvminor}),
 we conclude that {\it OF} is a molecular outflow
 and is part of the large-scale {\it HALO} outflow.

The  expansion velocity and the size of {\it OF} can be estimated
 from the averaged minor-axis $p-v$ diagram. 
In Figure~\ref{pvminor},
 the velocity range of {\it OF} is
about $\sim$ 180~km~s$^{-1}$ (Sec.~\ref{sec-pv}).
Since we do not see clear velocity gradient in a certain direction,
the simplest model for {\it OF} is an expanding outflow that
explodes isotropically at one side of the disk
(see the dashed curve in Figure~\ref{pvminor}).
The  expansion velocity and radius,
 which measured from the $p-v$ diagram,
 are $v_{\rm exp} \sim 90\pm10~{\rm km~s}^{-1}$ and
 $R_{\rm of} \sim~10\arcsec - 12\arcsec$  $\sim$ 370 -- 450~pc, respectively.

The molecular mass of {\it OF} can be estimated from CO flux.
The total flux, summed from each channel map, is
 82.0~Jy~km~s$^{-1}$ (= 1.14 $\times$ 10$^3$~K~km~s$^{-1}$).
Using Eq.~\ref{massh2} and Eq.~\ref{massmol},
 assuming the same conversion factor, 
 we can derive the gas mass of {\it OF},
 $M_{\rm of}$, as 2.8 $\times$ 10$^7$~M$_\sun$.
The mass is only a few percent of
 the total molecular gas detected in our data.

The mechanical energy is
$E_{\rm of,mech}$ = $\frac{1}{2}M_{\rm of}v_{\rm exp}^2$
 = (1.8 -- 2.8) $\times$ 10$^{54}$~erg.
After considering the mechanical efficiency $\gamma$,
 $\sim$ 10 -- 20 \%,  the energy transmitting from
supernova explosions into the surrounding ISM \citep{mcc87,wea77,lar74}.
The energy corresponds to 9,000 -- 28,000 supernova explosions.
Assuming the  expansion velocity keeps constant,
 the  expansion timescale is
$t_{\rm exp} = R_{\rm of} / v_{\rm exp}$ = 3.3 -- 6.8~Myr.
The average molecular gas mass flow rate can be calculated after
taking account the expansion timescale,
$\dot{M}_{\rm of} = M_{\rm of}/t_{\rm exp}$ 
$\sim$~4.1 -- 8.5~M$_\sun$~yr$^{-1}$.

\section{Discussion}
\subsection{Bar, Disk, or Superbubbles?}
\citet{irw96} claimed that they detected four molecular superbubbles
mostly associated with the low velocity gradient ridge outside of the nuclear disk.
The locations and the  expansion velocities of
 these four molecular superbubbles are
 shown in Figure~9 and Table~2 of \citet{irw96}.
Two superbubbles, A and C in \citet{irw96},
 are located at the same position
 and have the same central velocity but different  expansion velocities.
Another two superbubbles, B and D in \citet{irw96},
 are located at different positions
 and have different central velocities.

\begin{figure*}
\centering
\epsscale{1}
\plotone{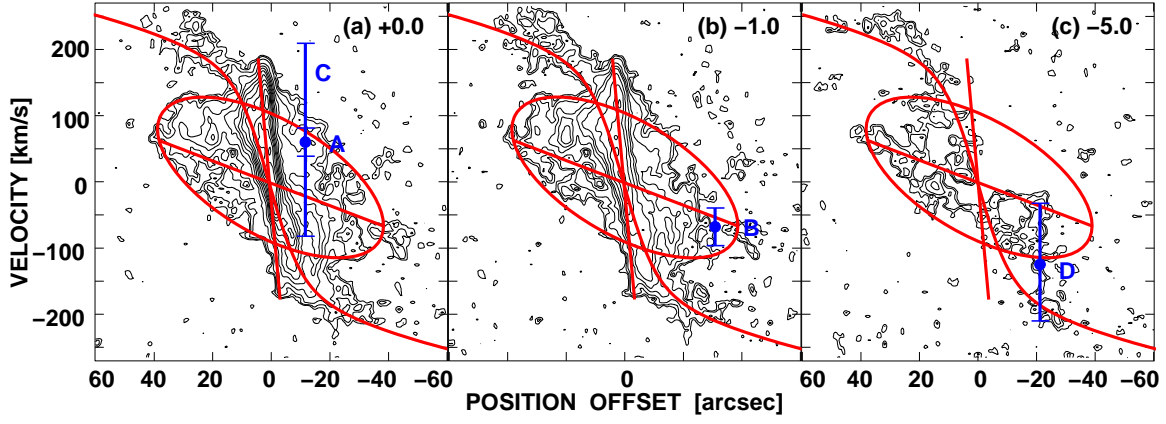}
\caption{
\label{abc-pvmajor}
Our NMA CO(1-0) $p-v$ diagrams
 marked with the position of superbubbles claimed by \citet{irw96}.
The red curve, solid lines, and ellipse 
 are the same with those in Figure~\ref{pvmajor0}.
The big blue dot in each $p-v$ diagram indicates the central velocity ($v_{\rm c}$)
 of the superbubble claimed by \citet{irw96}.
The blue bar indicates the  expansion velocity ($v_{\rm exp}$) of superbubbles.
The contour levels are
2, 3, 5, 10, 15, 20, ..., and 50$\sigma$,
where 1$\sigma$ = 9.09~mJy~beam$^{-1}$.
(a) The major-axis $p-v$ diagram.
Two superbubble A and C have the same central velocity of 55~km~s$^{-1}$.
Their  expansion velocities are 20~km~s$^{-1}$ and 150~km~s$^{-1}$, respectively.
(b) The $p-v$ diagram parallel to the major axis with offset of $-1.0\arcsec$.
The superbubble B has a central velocity of $-62$~km~s$^{-1}$
 and an  expansion velocity of 33~km~s$^{-1}$.
(c) The $p-v$ diagram parallel to the major axis with offset of $-5.0\arcsec$.
The superbubble D has a central velocity of $-126$~km~s$^{-1}$
 and an  expansion velocity of 85~km~s$^{-1}$.
}
\end{figure*}

\begin{figure*}
\centering
\epsscale{1}
\plotone{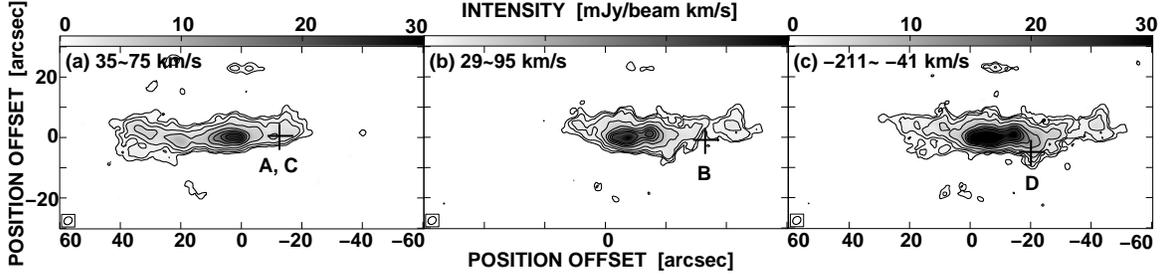}
\caption{
\label{abc-mnt0}
The integrated intensity maps.
The crosses mark the positions of superbubbles A, B, C, and D.
(a) Integrated velocity range between 35 and 75~km~s$^{-1}$.
The contour levels are 2, 5, 10, 20, 40, 60, 80, 100, 150, and 200$\sigma$,
 where 1$\sigma$ = 219.45~mJy~beam$^{-1}$~km~s$^{-1}$.
(b) Integrated velocity range between 29 and 95~km~s$^{-1}$.
The contour levels are 2, 5, 10, 20, 40, 60, 80, 100, and 150$\sigma$,
 where 1$\sigma$ = 194.75~mJy~beam$^{-1}$~km~s$^{-1}$.
(c) Integrated velocity range between $-211$ and $-41$~km~s$^{-1}$.
The contour levels are 2, 5, 10, 20, 40, 60, 80, and 100$\sigma$,
 where 1$\sigma$ = 229.70~mJy~beam$^{-1}$~km~s$^{-1}$.
}
\end{figure*}

In order to confirm these four structures are real superbubbles from our data,
 we used 
two types of images,
the $p-v$ diagrams (Figure~\ref{abc-pvmajor}) 
 and the intensity maps (Figure~\ref{abc-mnt0}),
 to  examine these structures.
We parametrized these structures with four parameters,
 ($\Delta\alpha$, $\Delta\delta$, $v_{\rm c}$, $v_{\rm exp}$),
 where $\Delta\alpha$ and $\Delta\delta$ specify the location,
 $v_{\rm c}$ is the central velocity of the superbubble,
 and $v_{\rm exp}$ is the  expansion velocity of the superbubble. 
The big blue dot in each $p-v$ diagram of Figure~\ref{abc-pvmajor} 
 is to mark the position and the central velocity of each superbubble.
The 
cross in each integrated intensity map of Figure~\ref{abc-mnt0}, 
 which is integrated along superbubble's  expansion velocity,
 is to mark the central location of the superbubble.

Superbubble A has a coordinate of ($-13\arcsec$, 0$\arcsec$, 55~km~s$^{-1}$, 20~km~s$^{-1}$).
In the $p-v$ diagram of Figure~\ref{abc-pvmajor}a, 
 we do not find any expanding structure at position A.
In addition, in the intensity map of Figure~\ref{abc-mnt0}a, 
 which is integrated along the velocity range of 35 -- 75~km~s$^{-1}$,
 we do not see any extended/distorted structure at position A.
Since it is difficult to match superbubble A at position A
 both in the $p-v$ diagram and the intensity map,
 it is difficult to conclude structure A as a superbubble.
Meanwhile, the position A in Figure~\ref{abc-pvmajor}a is close to the red ellipse,
 which represents the $p-v$ diagram of a bar structure.
This suggests that structure A is more likely part of the bar
 rather than a superbubble.

\begin{figure}
\centering
\epsscale{1}
\plotone{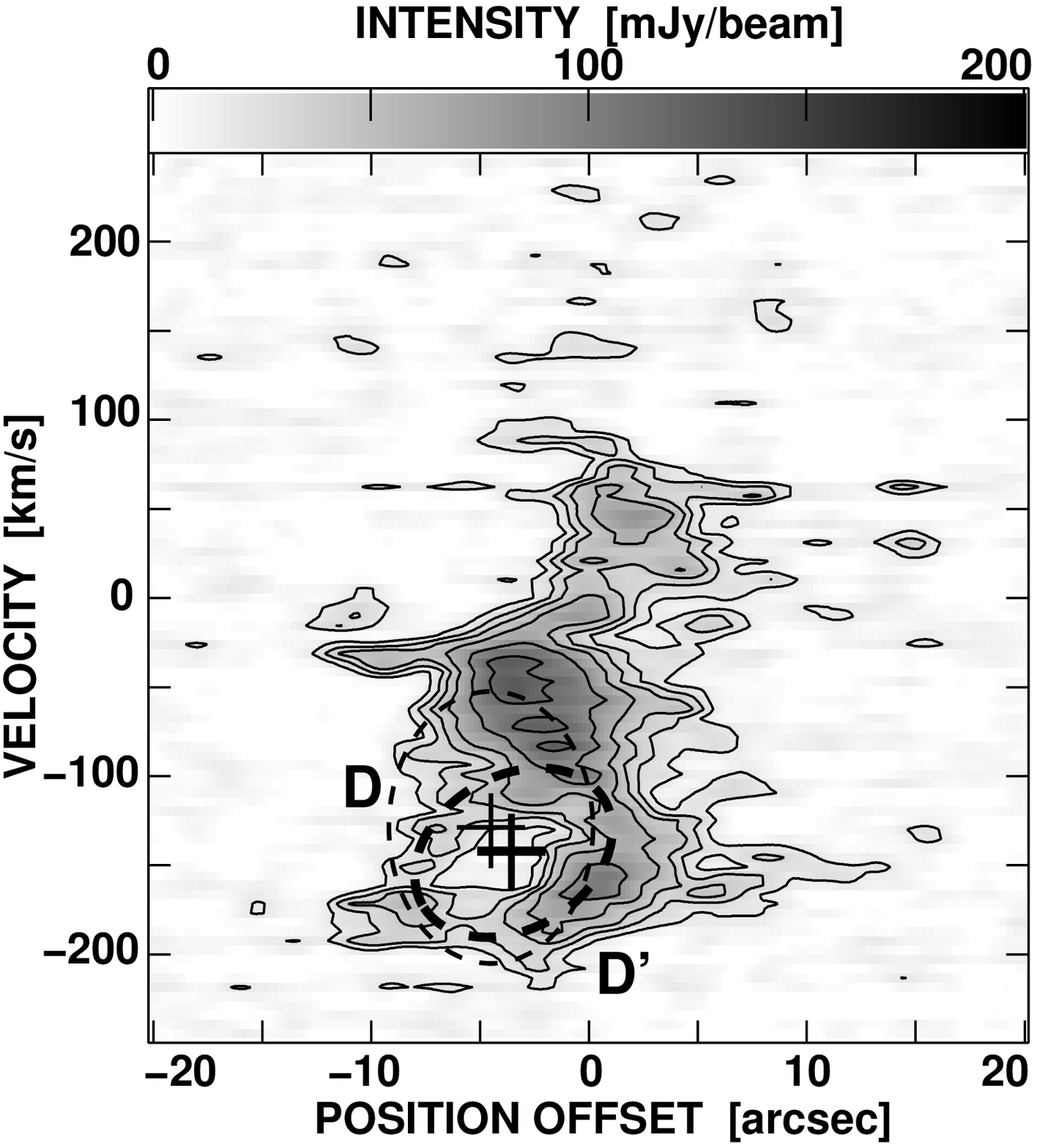}
\caption{
\label{abc-pvminor}
The minor-axis $p-v$ diagram of the superbubble D.
This is averaged between $-18\arcsec$ and $-22\arcsec$ along minor axis
 after rotating 17$\arcdeg$,
 the same rotating degree with that of \citet{irw96}.
The contour levels are 3, 5, 7, 10, 15, and 20$\sigma$,
 where 1$\sigma$ = 5.78~mJy beam$^{-1}$.
The thin cross and the thin dashed circle
 indicate the central position/velocity and the expansion velocity, respectively,
 of superbubble D identified by \citet{irw96}.
From our NMA data, we re-identify the superbubble as D'.
The thick dashed ellipse indicates the shell structure of superbubble D'.
The thick cross indicates the central position of superbubble D'.
}
\end{figure}

Superbubble B has a coordinate of ($-33\arcsec$, $-1\arcsec$, $-62$~km~s$^{-1}$, 33~km~s$^{-1}$).
Its $p-v$ diagram is shown in Figure~\ref{abc-pvmajor}b,
and its intensity map, which is integrated along the velocity range of 29 -- 95 km~s$^{-1}$,
 is shown in Figure~\ref{abc-mnt0}b.
Unfortunately, we do not find any expanding structure at position B in the $p-v$ diagram (Figure~\ref{abc-pvmajor}b),
neither find any shell structure at the position B in the intensity map (Figure~\ref{abc-mnt0}b).
Thus it is also difficult to identify structure B as a superbubble.
Besides, since position B in the $p-v$ diagram is located inside the red ellipse, 
 it is also possible to consider structure B as part of the bar.
On the other hand,
 the emission distribution at position B in the $p-v$ diagram 
 looks different from that at its symmetric position,
 which has a coordinate of ($33\arcsec$, $-1\arcsec$, $62$~km~s$^{-1}$, 33~km~s$^{-1}$).
Namely, there is almost no emission at position B,
 but there is emission detected at its symmetric position.
An explanation for this asymmetric is that
 the molecular gas at the position B has been blown up.
This indicates a possibility of the existence of a superbubble at position B in the past.
However, we can not distinguish which scenario is true.
Therefore, it is also difficult to conclude that structure B is a superbubble.

Superbubble C has a coordinate of ($-13\arcsec$, 0$\arcsec$, 55~km~s$^{-1}$, 150~km~s$^{-1}$),
 which has very similar coordinate with that of superbubble A.
 The only difference is their  expansion velocities.
Thus we can  examine superbubble C by
 using the same $p-v$ diagram with that of superbubble A
(Figure~\ref{abc-pvmajor}a). 
Again, we cannot find any emission represents superbubble C.
Thus it is still difficult to identify structure C as a superbubble.

Superbubble D has a coordinate 
 of ($-20\arcsec$, $-5\arcsec$, $-126$~km~s$^{-1}$, 85~km~s$^{-1}$).
We find emission detected at position D in the $p-v$ diagram of Figure~\ref{abc-pvmajor}c.
In addition, in the intensity map of Figure~\ref{abc-mnt0}c,
 which is integrated the velocity range between $-211$ and $-41$~km~s$^{-1}$,
 we find an extended structure at the position D.
We further check the $p-v$ diagram (Figure~\ref{abc-pvminor}) averaged
 between position offsets $-16\arcsec$ and $-24\arcsec$
 along the minor axis of Irwin \& Sofue's major axis
 (i.e., clockwise rotation of 17$\arcdeg$)
 in order to compare the two observations.
We  depict their superbubble D in Figure~\ref{abc-pvminor}
 by a  thin cross and a  thin dashed ellipse.
Interestingly, our deep observation detected more diffuse emission.
The enclosed loop marked as a  thick dashed ellipse,
 which represents the structure of a superbubble,
 is smaller than that of Irwin \& Sofue's  thin dashed loop.
 We identify the thick loop as the superbubble D' with a coordinate of
 ($-20\arcsec$, $-4\arcsec$, $-138$~km~s$^{-1}$, 50~km~s$^{-1}$).
Since a shift of 1$\arcsec$ from superbubble D
 is smaller than the  angular resolution of $\sim$ 3$\arcsec$,
 we conclude that the superbubble D and D' are identical,
but the superbubble D' has a more accurate coordinate than the superbubble D.
The radius of the superbubble D' can be estimated 
from Figure~\ref{abc-pvmajor}c and Figure~\ref{abc-pvminor},
 $\sim5\arcsec\times5\arcsec$ ($\sim$ 186~pc $\times$ 186~pc).

In conclusion, our results only confirm the existence of the superbubble D,
 superbubble A and C are likely to be parts of the bar structure,
 while superbubble B could be either part of the bar structure,
 or a past superbubble whose molecular gas has already been blown up.
The mis-interpretation is possibly
due to the low sensitivity of the previous observation.

\subsection{The Molecular Outflow is still Moving Outward?}
\label{n3628-pressure}
In order to figure out 
 the evolution of the molecular outflow {\it OF},
 we compare the pressure between the diffuse molecular gas {\it OF}
 and the plasma gas in the {\it CORE} region.

In the molecular outflow {\it OF}, 
the thermal pressure of constituent molecular cloud
 can be derived by $P_{\rm of,thm}$ = $n_{\rm of}kT$,
 where $n_{\rm of}$ is the number density of {\it OF}.
Since we do not have density and temperature information of {\it OF},
 we simply adopt typical values 
 (100 -- 1000~cm$^{-3}$ and 10 -- 100~K) for
 CO(1-0) emitting gas.
Thus the thermal pressure of {\it OF}
 is 1.4 $\times$ 10$^{-(11-13)}$~dyne~cm$^{-2}$.

In {\it CORE} region,
 the thermal pressures of the plasma outflow
 in the lower and the higher temperatures are
1.4$f^{-1/2}$ $\times$ 10$^{-11}$~dyne~cm$^{-2}$ and
 4.8$f^{-1/2}$ $\times$ 10$^{-10}$~dyne~cm$^{-2}$, respectively 
(see Sec.~\ref{sec-ion-prop} and Tab.~\ref{xray-prop}).
Adopting a volume filling factor of 0.1 -- 0.01, 
 their thermal pressures are 
(0.5 -- 1.4) $\times$ 10$^{-10}$~dyne~cm$^{-2}$ and
(1.5 -- 4.8) $\times$ 10$^{-9}$~dyne~cm$^{-2}$, respectively.
Besides, 
the ram pressure that {\it OF} suffered from the plasma outflow
 can be derived by
$P_{\rm ram}$ = $\rho_{\rm plm} \times v_{\rm rel}^2$,
 where $v_{\rm rel}=|v_{\rm exp}-v_{\rm plm}|$ 
 is the relative velocity between
 the  expansion velocity of {\it OF}, $v_{\rm exp}$,
 and of the plasma outflow, $v_{\rm plm}$.
We assume that the  expansion velocity of the plasma outflow
 is similar to its thermal velocity,
i.e., $v_{\rm plm}\sim\sqrt{3kT/\mu m_{\rm p}}$. 
Thus $v_{\rm plm}$ estimated from the lower and the higher temperature components in the {\it CORE} are
 520 and 2,050~km~s$^{-1}$, respectively.
Since $v_{\rm exp}\sim$ 90~km~s$^{-1}$,
 $v_{\rm rel}$ for the lower and the higher temperature components
in the {\it CORE} are
 430 and 1,960~km~s$^{-1}$, respectively,
then the ram pressures of the lower and the higher temperature components are
 2.9~$f^{-1/2}$ $\times$ 10$^{-11}$~dyne~cm$^{-2}$ and
 1.3~$f^{-1/2}$ $\times$ 10$^{-9}$~dyne~cm$^{-2}$, respectively.
After considering a filling factor of 0.01 -- 0.1, 
 the ram pressures are 
(0.9 -- 3.1) $\times$ 10$^{-10}$~dyne~cm$^{-2}$ and
(0.4 -- 1.3) $\times$ 10$^{-8}$~dyne~cm$^{-2}$, respectively. 
We notice that in the lower temperature components,
the ram pressure and the plasma thermal pressure have similar values,
while in the higher temperature components,
the ram pressure is higher than the plasma thermal pressure.
Generally speaking, the ram pressures 
and the plasma thermal pressures
 in {\it CORE}
are $\sim$ 10$^{-(8-10)}$~dyne~cm$^{-2}$.
Since the thermal pressure of the molecular outflow {\it OF} is
$\sim$ 10$^{-(11-13)}$ ~dyne~cm$^{-2}$ (Sec.~\ref{n3628-pressure}),
this indicates that the {\it OF} is pushed by the plasma gas.
Thus we conclude that the {\it OF} is still moving outward.

\subsection{The Evolutionary Stage of the Starburst Activity}
In order to figure out the evolutionary stage of the starburst activity,
we defined an evolutionary parameter, $\tau$,
 which is the  expansion timescale of molecular outflow, $t_{\rm exp}$,
normalized to the total starburst timescale, $t_{\rm SB}$,
\begin{eqnarray}
\label{eq-tau}
\tau &=& \frac{t_{\rm exp}}{t_{\rm SB}}, \\
 &=& \frac{t_{\rm exp}}{t_{\rm exp}+t_{\rm cons}}, 
\end{eqnarray}
where $\tau$ = 1 means the starburst period is finished,
and 
 $t_{\rm SB}$ 
 includes two timescales,
$t_{\rm exp}$,
 implying the timescale from the beginning of the starburst to now,
and $t_{\rm cons}$, the molecular gas consumption timescale, 
 i.e., the timescale that the current molecular gas
 in the starburst region to be used up,
 implying the timescale from now to the end of starburst period.
$t_{\rm cons}$ is highly depending on two processes,
 mass loss through molecular outflow, and star formation,
\begin{eqnarray}
\label{eq-csump}
t_{\rm cons} &=& \frac{M_{\rm SBR}}{\dot{M}_{\rm of}+{\rm SFR}}, 
\end{eqnarray}
where $M_{\rm SBR}$ is the molecular gas mass in the starburst region,
 and the SFR is the star formation rate.

From Sec.~\ref{n3628-mol-prop}, we know the mass loss rate,
$\dot{M}_{\rm of}$ = 4.1 -- 8.5~M$_{\sun}$~yr$^{-1}$.
We define the size of the starburst region as the same with the base size of the X-ray emission,
 which is $\sim$~0.5~kpc.
Since the CO flux within starburst region is $\sim$~450~Jy~km~s$^{-1}$,
this leads the molecular gas mass to be
 $\sim$~2.0$\times$10$^{8}$~M$_{\sun}$.
Since our data do not provide any information of the SFR, 
 we use IR luminosity, L$_{\rm IR}$, to estimate the SFR as the upper limit.
We assume that the IR luminosity is all contributed from the central starburst region,
 thus the SFR in the starburst region can be estimated as,
 SFR = $4.5\times10^{-44}\ L_{\rm IR}$ \citep{ken98}.
Taking $L_{\rm IR}$ = $10^{10.25}$ L$_{\sun}$ \citep{san03},
 SFR is 3.2~M$_{\sun}$~yr$^{-1}$.
Thus $t_{\rm cons}$ of NGC~3628 is calculated as 17 -- 27~Myr.

Adopting a molecular outflow  expansion timescale of 3.3 -- 6.8~Myr
 (from Sec.~\ref{n3628-mol-prop}),
 we have $t_{\rm SB}$ = 20 -- 34~Myr, and $\tau$ = 0.11 -- 0.25.
This suggests that the evolutionary stage of the starburst activity is relatively young.

\section{Summary}
Our NMA CO(1-0) observations detected
a sub-kpc $\sim$ 370 -- 450~pc scale molecular outflow for the first time
with an  expansion speed of $\sim90\pm10$~km~s$^{-1}$
in the starburst galaxy NGC~3628.
The molecular outflow mass is 2.8~$\times$~10$^7$~M$_{\sun}$,
 the  expansion timescale is 3.3 -- 6.8~Myr,
 and the mechanical energy is (1.8 -- 2.8)~$\times$~10$^{54}$~erg.

In order to understand the evolution of molecular outflow
 and the relation with the plasma outflow,
 we compared our NMA CO(1-0) data with the CXO X-ray archival data.
We estimated the total energy both from mechanical energy of molecular outflow
and thermal energy of plasma outflow.
The total energy generated by the starburst activity is estimated as
 (2.3 -- 2.8)~$\times$ 10$^{55}$~erg.
The thermal pressure  of the molecular outflow is 
 $\sim$ 10$^{-(11-13)}$~dyne~cm$^{-2}$. 
The thermal pressure and the ram pressure of the plasma outflow are
 $\sim$ 10$^{-(8-10)}$~dyne~cm$^{-2}$.
This indicates that 
the molecular outflow is still pushed by the plasma gas and 
 is moving outward.
 Besides, 
 we confirmed one superbubble as a real structure
 from the four superbubbles that \citet{irw96} claimed.
 Two of them should be just noise, 
 and the other one is difficult to be concluded.

We also estimated the molecular gas consumption timescale, 
 17 -- 27~Myr.
This yields a total starburst timescale to be 20 -- 34~Myr
with an evolutionary parameter of 0.11 -- 0.25,
 suggesting that the starburst activity in NGC 3628
 is still in a young stage.

\bigskip
\begin{acknowledgements}
We appreciate to Yi-Jung Yang, I-Non Chiu, and Dr. Sandor Molnar
for exchanging ideas of X-ray data analysis,
and to Dr. Kazushi Sakamoto and Prof. Mousumi Das Amarnath 
for several suggestion on this paper.
We also thank the anonymous referee for very useful comments.
We are grateful to the NRO staff for the operation and improvement of
the NMA.
This work is supported by the National Science Council (NSC) of Taiwan, 
NSC  97-2112-M-001-021-MY3 and  NSC 100-2112-M-001-006-MY3.
\end{acknowledgements}

\begin{deluxetable}{lllcccccccccccccc}
\tabletypesize{\scriptsize}
\tablecaption{Spectral Models for the CXO Spectra
\label{xray-fit}}
\tablewidth{0pt}
\tablehead{
\colhead{Region} &
\colhead{$N_{\rm H}$} &
\colhead{$kT_1$} &
\colhead{Norm$_1$} &
\colhead{$kT_2$} &
\colhead{Norm$_2$} &
\colhead{O} &
\colhead{Mg} &
\colhead{Si} &
\colhead{Fe} &
\colhead{Red.$\chi^2$} &
\colhead{D.O.F.}
\\
\colhead{} &
\colhead{(1)} &
\colhead{(2)} &
\colhead{(3)} &
\colhead{(4)} &
\colhead{(5)} &
\colhead{(6)} &
\colhead{(7)} &
\colhead{(8)} &
\colhead{(9)} &
\colhead{(10)} &
\colhead{(11)}
}
\startdata
HALO\tablenotemark{a} &
0.02\tablenotemark{*} &
0.21$_{-0.02}^{+0.03}$ &
8.52$_{-0.05}^{+0.04}$ &
0.60$_{-0.15}^{+0.57}$ &
4.63$_{-2.58}^{+2.60}$ &
0.48$_{-0.14}^{+0.24}$ &
1\tablenotemark{*} &
2.80$_{-1.81}^{+2.26}$ &
0.39$_{-0.13}^{+0.24}$ &
1.25 &
96
\\
\hline
CORE\tablenotemark{b} &
0.33$_{-0.20}^{+0.33}$ &
0.47$_{-0.17}^{+0.15}$ &
0.93$_{-0.80}^{+3.73}$ &
7.29$_{-3.42}^{+28.0}$ &
2.18$_{-0.64}^{+0.71}$ &
1\tablenotemark{*} &
5.0$_{-3.5}^{+20.9}$ &
13.5$_{-8.2}^{+37.3}$ &
0.81$_{-0.49}^{+0.71}$ &
1.56 &
14
\enddata
\tablecomments
{
(1) Foreground absorbing column density of the VMEKAL in units of 10$^{22}$~cm$^{-2}$.
(2) Temperature of component 1 in units of keV,
where 1 represents the lower temperature component. 
(3) Normalization of component 1 in units of
$10^{-5}\int 10^{-14}n_{\rm e}n_{\rm H}dV/4\pi D^2$,
where $n_{\rm e}$ is the electron number density,
$n_{\rm H}$ is the hydrogen number density,
$V$ is the volume of plasma, and
$D$ is the distance between the sources and the observer.
(4) Temperature of component 2 in units of keV,
where 2 represent the higher temperature component.
(5) Normalization of component 2,
 the units is the same with that of component 1.
(6) O abundance in units of $Z_\sun$.
(7) Mg abundance in units of $Z_\sun$.
(8) Si abundance in units of $Z_\sun$.
(9) Fe abundance in units of $Z_\sun$.
(10) Reduced $\chi^{2}$ of fit.
(11) Degrees of freedom in the fit.
}
\tablenotetext{a}{Fitted spectrum in the energy range 0.4 -- 3.0~keV.}
\tablenotetext{b}{Fitted spectrum in the energy range 0.3 -- 7.0~keV.}
\tablenotetext{*}{Denotes parameters held fixed.}
\end{deluxetable}

\begin{deluxetable}{ccccrrrrcccccc}
\tabletypesize{\scriptsize}
\tablecaption{The properties of plasma outflow in NGC 3628
\label{xray-prop}}
\tablewidth{0pt}
\tablehead{
\colhead{Region} &
\colhead{$R$} &
\colhead{Component} &
\colhead{$kT$} &
\colhead{$EI$} &
\colhead{$n_{\rm e}$} &
\colhead{$M_{\rm plm}$} &
\colhead{$P_{\rm plm,thm}$} &
\colhead{$E_{\rm plm,thm}$}
\\
\colhead{} &
\colhead{(1)} &
\colhead{(2)} &
\colhead{(3)} &
\colhead{(4)} &
\colhead{(5)} &
\colhead{(6)} &
\colhead{(7)} &
\colhead{(8)}
}
\startdata
HALO &
6.72 &
1 &
0.21 &
6.07 &
0.13 $f^{-1/2}$ &
40.0 $f^{1/2}$ &
0.09 $f^{-1/2}$ &
4.80 $f^{1/2}$
\\
 &
 &
2 &
0.60 &
3.30 &
0.09 $f^{-1/2}$ &
29.4 $f^{1/2}$ &
0.18 $f^{-1/2}$ &
10.1 $f^{1/2}$
\\
\hline
CORE&
0.67 &
1 &
0.47 &
0.34 &
0.95 $f^{-1/2}$ &
0.30 $f^{1/2}$ &
1.43 $f^{-1/2}$ &
0.08 $f^{1/2}$
\\
 &
 &
2 &
7.29 &
1.55 &
2.04 $f^{-1/2}$ &
0.64 $f^{1/2}$ &
47.7 $f^{-1/2}$ &
2.67 $f^{1/2}$
\enddata
\tablecomments{
(1) The radius of outflow in units of kpc.
(2) The temperature component in absorbed two-temperature model, WABS*(VMEKAL + VMEKAL), 
where 1 represents the lower temperature component, 
and 2 represent the higher temperature component.
(3) Temperature in units of keV.
(4) Emission Integral ($EI = \int n_{e}n_{H}dV$) in units of $10^{61}~cm^{-3}$.
(5) Number density in units of 10$^{-2}$~cm$^{-3}$,
 where $f$ is the filling factor.
(6) Mass in units of $10^6$ M$_\sun$.
(7) Thermal pressure in units of 10$^{-11}$~dyne~cm$^{-2}$.
(8) Thermal energy in units of 10$^{55}$~erg.
}
\end{deluxetable}

\end{document}